%
%
%
%
%
%
%
%
\def\standardrisposta{s }\def\reducedrisposta{r }
\def\mplarisposta{mpla }\def\zerorisposta{z }
\def\doublerisposta{d }\def\cartarisposta{e }\def\amsrisposta{y }
\newcount\ingrandimento \newcount\sinnota \newcount\dimnota
\newcount\unoduecol \newdimen\collhsize \newdimen\tothsize
\newdimen\fullhsize \newcount\controllorisposta \sinnota=1
\newskip\infralinea  \global\controllorisposta=0
\immediate\write16 { ********  Welcome to PANDA macros (Plain TeX,
AP, 1991) ******** }
%
%
%
\def\risposta{s } 
\def\srisposta{e }
\def\arisposta{y }
\ifx\risposta\standardrisposta \ingrandimento=1200
\message {>> This will come out UNREDUCED << }
\dimnota=2 \unoduecol=1 \global\controllorisposta=1 \fi
\ifx\risposta\reducedrisposta \ingrandimento=1095 \dimnota=1
\unoduecol=1  \global\controllorisposta=1
\message {>> This will come out REDUCED << } \fi
\ifx\risposta\doublerisposta \ingrandimento=1000 \dimnota=2
\unoduecol=2   \message {>> You must print this in
LANDSCAPE orientation << } \global\controllorisposta=1 \fi
\ifx\risposta\mplarisposta \ingrandimento=1000 \dimnota=1
\message {>> Mod. Phys. Lett. A format << }
\unoduecol=1 \global\controllorisposta=1 \fi
\ifx\risposta\zerorisposta \ingrandimento=1000 \dimnota=2
\message {>> Zero Magnification format << }
\unoduecol=1 \global\controllorisposta=1 \fi
\ifnum\controllorisposta=0  \ingrandimento=1200
\message {>>> ERROR IN INPUT, I ASSUME STANDARD
UNREDUCED FORMAT <<< }  \dimnota=2 \unoduecol=1 \fi
\magnification=\ingrandimento
%
%
%
%
\newdimen\eucolumnsize \newdimen\eudoublehsize \newdimen\eudoublevsize
\newdimen\uscolumnsize \newdimen\usdoublehsize \newdimen\usdoublevsize
\newdimen\eusinglehsize \newdimen\eusinglevsize \newdimen\ussinglehsize
\newskip\standardbaselineskip \newdimen\ussinglevsize
\newskip\reducedbaselineskip \newskip\doublebaselineskip
\eucolumnsize=12.0truecm    
\eudoublehsize=25.5truecm   
\eudoublevsize=6.7truein    
\uscolumnsize=4.4truein     
\usdoublehsize=9.4truein    
\usdoublevsize=6.8truein    
\eusinglehsize=6.5truein    
\eusinglevsize=24truecm     
\ussinglehsize=6.5truein    
\ussinglevsize=8.9truein    
\standardbaselineskip=16pt plus.2pt  
\reducedbaselineskip=14pt plus.2pt   
\doublebaselineskip=12pt plus.2pt    
%
%
\def\Portoffset{}
\def\Landoffset{\voffset=-.2truein}
\ifx\risposta\mplarisposta \def\Portoffset{\hoffset=1.8truecm} \fi
%
%
\def\Landspec{}
\tolerance=10000
\parskip=0pt plus2pt  \leftskip=0pt \rightskip=0pt
%
%
\ifx\risposta\standardrisposta \infralinea=\standardbaselineskip \fi
\ifx\risposta\reducedrisposta  \infralinea=\reducedbaselineskip \fi
\ifx\risposta\doublerisposta   \infralinea=\doublebaselineskip \fi
\ifx\risposta\mplarisposta     \infralinea=13pt \fi
\ifx\risposta\zerorisposta     \infralinea=12pt plus.2pt\fi
\ifnum\controllorisposta=0    \infralinea=\standardbaselineskip \fi
\ifx\risposta\doublerisposta   \Landoffset \else \Portoffset \fi
\ifx\risposta\doublerisposta \ifx\srisposta\cartarisposta
\tothsize=\eudoublehsize \collhsize=\eucolumnsize
\vsize=\eudoublevsize  \else  \tothsize=\usdoublehsize
\collhsize=\uscolumnsize \vsize=\usdoublevsize \fi \else
\ifx\srisposta\cartarisposta \tothsize=\eusinglehsize
\vsize=\eusinglevsize \else  \tothsize=\ussinglehsize
\vsize=\ussinglevsize \fi \collhsize=4.4truein \fi
\ifx\risposta\mplarisposta \tothsize=5.0truein
\vsize=7.8truein \collhsize=4.4truein \fi
%
%
%
%
\newcount\contaeuler \newcount\contacyrill \newcount\contaams
\font\ninerm=cmr9  \font\eightrm=cmr8  \font\sixrm=cmr6
\font\ninei=cmmi9  \font\eighti=cmmi8  \font\sixi=cmmi6
\font\ninesy=cmsy9  \font\eightsy=cmsy8  \font\sixsy=cmsy6
\font\ninebf=cmbx9  \font\eightbf=cmbx8  \font\sixbf=cmbx6
\font\ninett=cmtt9  \font\eighttt=cmtt8  \font\nineit=cmti9
\font\eightit=cmti8 \font\ninesl=cmsl9  \font\eightsl=cmsl8
\skewchar\ninei='177 \skewchar\eighti='177 \skewchar\sixi='177
\skewchar\ninesy='60 \skewchar\eightsy='60 \skewchar\sixsy='60
\hyphenchar\ninett=-1 \hyphenchar\eighttt=-1 \hyphenchar\tentt=-1
\def\bfmath{\cmmib}                 
\font\tencmmib=cmmib10  \newfam\cmmibfam  \skewchar\tencmmib='177
\font\tencmbsy=cmbsy10  \newfam\cmbsyfam  \skewchar\tencmbsy='60
\def\scaps{\cmcsc}                 
\font\tencmcsc=cmcsc10  \newfam\cmcscfam
\ifnum\ingrandimento=1095

\font\capsone=cmcsc10 at 10.95pt 

\else

\font\capsone=cmcsc10 at 12pt 
\fi

\def\ttaarr{\bf}		
\def\ppaarr{\sl}		

%
%
%
\newfam\eufmfam \newfam\msamfam \newfam\msbmfam \newfam\eufbfam
\def\Loadeulerfonts{\global\contaeuler=1 \ifx\arisposta\amsrisposta
\font\teneufm=eufm10              
\font\eighteufm=eufm8 \font\nineeufm=eufm9 \font\sixeufm=eufm6
\font\seveneufm=eufm7  \font\fiveeufm=eufm5
\font\teneufb=eufb10              
\font\eighteufb=eufb8 \font\nineeufb=eufb9 \font\sixeufb=eufb6
\font\seveneufb=eufb7  \font\fiveeufb=eufb5
\font\teneurm=eurm10              
\font\eighteurm=eurm8 \font\nineeurm=eurm9
\font\teneurb=eurb10              
\font\eighteurb=eurb8 \font\nineeurb=eurb9
\font\teneusm=eusm10              
\font\eighteusm=eusm8 \font\nineeusm=eusm9
\font\teneusb=eusb10              
\font\eighteusb=eusb8 \font\nineeusb=eusb9
\else \def\eufm{\tt} \def\eufb{\tt} \def\eurm{\tt} \def\eurb{\tt}
\def\eusm{\tt} \def\eusb{\tt}    \fi}
\def\loadeuler{\Loadeulerfonts\tenpoint}
\def\loadamsmath{\global\contaams=1 \ifx\arisposta\amsrisposta
\font\tenmsam=msam10 \font\ninemsam=msam9 \font\eightmsam=msam8
\font\sevenmsam=msam7 \font\sixmsam=msam6 \font\fivemsam=msam5
\font\tenmsbm=msbm10 \font\ninemsbm=msbm9 \font\eightmsbm=msbm8
\font\sevenmsbm=msbm7 \font\sixmsbm=msbm6 \font\fivemsbm=msbm5
\else \def\msbm{\bf} \fi \def\Bbb{\msbm} \def\symbl{\msam} \tenpoint}
\def\loadcyrill{\global\contacyrill=1 \ifx\arisposta\amsrisposta
\font\tenwncyr=wncyr10 \font\ninewncyr=wncyr9 \font\eightwncyr=wncyr8
\font\tenwncyb=wncyr10 \font\ninewncyb=wncyr9 \font\eightwncyb=wncyr8
\font\tenwncyi=wncyr10 \font\ninewncyi=wncyr9 \font\eightwncyi=wncyr8
\else \def\cyrill{\sl} \def\cyrilb{\sl} \def\cyrili{\sl} \fi\tenpoint}
\ifx\arisposta\amsrisposta
\font\sevenex=cmex7               
\font\eightex=cmex8  \font\nineex=cmex9
\font\ninecmmib=cmmib9   \font\eightcmmib=cmmib8
\font\sevencmmib=cmmib7 \font\sixcmmib=cmmib6
\font\fivecmmib=cmmib5   \skewchar\ninecmmib='177
\skewchar\eightcmmib='177  \skewchar\sevencmmib='177
\skewchar\sixcmmib='177   \skewchar\fivecmmib='177
%
%
%
\def\ninecmbsy{\tencmbsy}
\def\eightcmbsy{\tencmbsy}
\def\sevencmbsy{\tencmbsy}
\def\sixcmbsy{\tencmbsy}
\def\fivecmbsy{\tencmbsy}
\font\ninecmcsc=cmcsc9    \font\eightcmcsc=cmcsc8     \else
\def\cmmib{\fam\cmmibfam\tencmmib}\textfont\cmmibfam=\tencmmib
\scriptfont\cmmibfam=\tencmmib \scriptscriptfont\cmmibfam=\tencmmib
\def\cmbsy{\fam\cmbsyfam\tencmbsy} \textfont\cmbsyfam=\tencmbsy
\scriptfont\cmbsyfam=\tencmbsy \scriptscriptfont\cmbsyfam=\tencmbsy
\scriptfont\cmcscfam=\tencmcsc \scriptscriptfont\cmcscfam=\tencmcsc
\def\cmcsc{\fam\cmcscfam\tencmcsc} \textfont\cmcscfam=\tencmcsc \fi
\catcode`@=11
\newskip\ttglue
\gdef\tenpoint{\def\rm{\fam0\tenrm}
  \textfont0=\tenrm \scriptfont0=\sevenrm \scriptscriptfont0=\fiverm
  \textfont1=\teni \scriptfont1=\seveni \scriptscriptfont1=\fivei
  \textfont2=\tensy \scriptfont2=\sevensy \scriptscriptfont2=\fivesy
  \textfont3=\tenex \scriptfont3=\tenex \scriptscriptfont3=\tenex
  \def\mcal{\fam2 \tensy}  \def\mmit{\fam1 \teni}
  \textfont\itfam=\tenit \def\it{\fam\itfam\tenit}
  \textfont\slfam=\tensl \def\sl{\fam\slfam\tensl}
  \textfont\ttfam=\tentt \scriptfont\ttfam=\eighttt
  \scriptscriptfont\ttfam=\eighttt  \def\tt{\fam\ttfam\tentt}
  \textfont\bffam=\tenbf \scriptfont\bffam=\sevenbf
  \scriptscriptfont\bffam=\fivebf \def\bf{\fam\bffam\tenbf}
     \ifx\arisposta\amsrisposta    \ifnum\contaeuler=1
  \textfont\eufmfam=\teneufm \scriptfont\eufmfam=\seveneufm
  \scriptscriptfont\eufmfam=\fiveeufm \def\eufm{\fam\eufmfam\teneufm}
  \textfont\eufbfam=\teneufb \scriptfont\eufbfam=\seveneufb
  \scriptscriptfont\eufbfam=\fiveeufb \def\eufb{\fam\eufbfam\teneufb}
  \def\eurm{\teneurm} \def\eurb{\teneurb} \def\eusm{\teneusm}
  \def\eusb{\teneusb}    \fi    \ifnum\contaams=1
  \textfont\msamfam=\tenmsam \scriptfont\msamfam=\sevenmsam
  \scriptscriptfont\msamfam=\fivemsam \def\msam{\fam\msamfam\tenmsam}
  \textfont\msbmfam=\tenmsbm \scriptfont\msbmfam=\sevenmsbm
  \scriptscriptfont\msbmfam=\fivemsbm \def\msbm{\fam\msbmfam\tenmsbm}
     \fi      \ifnum\contacyrill=1     \def\cyrill{\tenwncyr}
  \def\cyrilb{\tenwncyb}  \def\cyrili{\tenwncyi}         \fi
  \textfont3=\tenex \scriptfont3=\sevenex \scriptscriptfont3=\sevenex
  \def\cmmib{\fam\cmmibfam\tencmmib} \scriptfont\cmmibfam=\sevencmmib
  \textfont\cmmibfam=\tencmmib  \scriptscriptfont\cmmibfam=\fivecmmib
  \def\cmbsy{\fam\cmbsyfam\tencmbsy} \scriptfont\cmbsyfam=\sevencmbsy
  \textfont\cmbsyfam=\tencmbsy  \scriptscriptfont\cmbsyfam=\fivecmbsy
  \def\cmcsc{\fam\cmcscfam\tencmcsc} \scriptfont\cmcscfam=\eightcmcsc
  \textfont\cmcscfam=\tencmcsc \scriptscriptfont\cmcscfam=\eightcmcsc
     \fi            \tt \ttglue=.5em plus.25em minus.15em
  \normalbaselineskip=12pt
  \setbox\strutbox=\hbox{\vrule height8.5pt depth3.5pt width0pt}
  \let\sc=\eightrm \let\big=\tenbig   \normalbaselines
  \baselineskip=\infralinea  \rm}
\gdef\ninepoint{\def\rm{\fam0\ninerm}
  \textfont0=\ninerm \scriptfont0=\sixrm \scriptscriptfont0=\fiverm
  \textfont1=\ninei \scriptfont1=\sixi \scriptscriptfont1=\fivei
  \textfont2=\ninesy \scriptfont2=\sixsy \scriptscriptfont2=\fivesy
  \textfont3=\tenex \scriptfont3=\tenex \scriptscriptfont3=\tenex
  \def\mcal{\fam2 \ninesy}  \def\mmit{\fam1 \ninei}
  \textfont\itfam=\nineit \def\it{\fam\itfam\nineit}
  \textfont\slfam=\ninesl \def\sl{\fam\slfam\ninesl}
  \textfont\ttfam=\ninett \scriptfont\ttfam=\eighttt
  \scriptscriptfont\ttfam=\eighttt \def\tt{\fam\ttfam\ninett}
  \textfont\bffam=\ninebf \scriptfont\bffam=\sixbf
  \scriptscriptfont\bffam=\fivebf \def\bf{\fam\bffam\ninebf}
     \ifx\arisposta\amsrisposta  \ifnum\contaeuler=1
  \textfont\eufmfam=\nineeufm \scriptfont\eufmfam=\sixeufm
  \scriptscriptfont\eufmfam=\fiveeufm \def\eufm{\fam\eufmfam\nineeufm}
  \textfont\eufbfam=\nineeufb \scriptfont\eufbfam=\sixeufb
  \scriptscriptfont\eufbfam=\fiveeufb \def\eufb{\fam\eufbfam\nineeufb}
  \def\eurm{\nineeurm} \def\eurb{\nineeurb} \def\eusm{\nineeusm}
  \def\eusb{\nineeusb}     \fi   \ifnum\contaams=1
  \textfont\msamfam=\ninemsam \scriptfont\msamfam=\sixmsam
  \scriptscriptfont\msamfam=\fivemsam \def\msam{\fam\msamfam\ninemsam}
  \textfont\msbmfam=\ninemsbm \scriptfont\msbmfam=\sixmsbm
  \scriptscriptfont\msbmfam=\fivemsbm \def\msbm{\fam\msbmfam\ninemsbm}
     \fi       \ifnum\contacyrill=1     \def\cyrill{\ninewncyr}
  \def\cyrilb{\ninewncyb}  \def\cyrili{\ninewncyi}         \fi
  \textfont3=\nineex \scriptfont3=\sevenex \scriptscriptfont3=\sevenex
  \def\cmmib{\fam\cmmibfam\ninecmmib}  \textfont\cmmibfam=\ninecmmib
  \scriptfont\cmmibfam=\sixcmmib \scriptscriptfont\cmmibfam=\fivecmmib
  \def\cmbsy{\fam\cmbsyfam\ninecmbsy}  \textfont\cmbsyfam=\ninecmbsy
  \scriptfont\cmbsyfam=\sixcmbsy \scriptscriptfont\cmbsyfam=\fivecmbsy
  \def\cmcsc{\fam\cmcscfam\ninecmcsc} \scriptfont\cmcscfam=\eightcmcsc
  \textfont\cmcscfam=\ninecmcsc \scriptscriptfont\cmcscfam=\eightcmcsc
     \fi            \tt \ttglue=.5em plus.25em minus.15em
  \normalbaselineskip=11pt
  \setbox\strutbox=\hbox{\vrule height8pt depth3pt width0pt}
  \let\sc=\sevenrm \let\big=\ninebig \normalbaselines\rm}
\gdef\eightpoint{\def\rm{\fam0\eightrm}
  \textfont0=\eightrm \scriptfont0=\sixrm \scriptscriptfont0=\fiverm
  \textfont1=\eighti \scriptfont1=\sixi \scriptscriptfont1=\fivei
  \textfont2=\eightsy \scriptfont2=\sixsy \scriptscriptfont2=\fivesy
  \textfont3=\tenex \scriptfont3=\tenex \scriptscriptfont3=\tenex
  \def\mcal{\fam2 \eightsy}  \def\mmit{\fam1 \eighti}
  \textfont\itfam=\eightit \def\it{\fam\itfam\eightit}
  \textfont\slfam=\eightsl \def\sl{\fam\slfam\eightsl}
  \textfont\ttfam=\eighttt \scriptfont\ttfam=\eighttt
  \scriptscriptfont\ttfam=\eighttt \def\tt{\fam\ttfam\eighttt}
  \textfont\bffam=\eightbf \scriptfont\bffam=\sixbf
  \scriptscriptfont\bffam=\fivebf \def\bf{\fam\bffam\eightbf}
     \ifx\arisposta\amsrisposta   \ifnum\contaeuler=1
  \textfont\eufmfam=\eighteufm \scriptfont\eufmfam=\sixeufm
  \scriptscriptfont\eufmfam=\fiveeufm \def\eufm{\fam\eufmfam\eighteufm}
  \textfont\eufbfam=\eighteufb \scriptfont\eufbfam=\sixeufb
  \scriptscriptfont\eufbfam=\fiveeufb \def\eufb{\fam\eufbfam\eighteufb}
  \def\eurm{\eighteurm} \def\eurb{\eighteurb} \def\eusm{\eighteusm}
  \def\eusb{\eighteusb}       \fi    \ifnum\contaams=1
  \textfont\msamfam=\eightmsam \scriptfont\msamfam=\sixmsam
  \scriptscriptfont\msamfam=\fivemsam \def\msam{\fam\msamfam\eightmsam}
  \textfont\msbmfam=\eightmsbm \scriptfont\msbmfam=\sixmsbm
  \scriptscriptfont\msbmfam=\fivemsbm \def\msbm{\fam\msbmfam\eightmsbm}
     \fi       \ifnum\contacyrill=1     \def\cyrill{\eightwncyr}
  \def\cyrilb{\eightwncyb}  \def\cyrili{\eightwncyi}         \fi
  \textfont3=\eightex \scriptfont3=\sevenex \scriptscriptfont3=\sevenex
  \def\cmmib{\fam\cmmibfam\eightcmmib}  \textfont\cmmibfam=\eightcmmib
  \scriptfont\cmmibfam=\sixcmmib \scriptscriptfont\cmmibfam=\fivecmmib
  \def\cmbsy{\fam\cmbsyfam\eightcmbsy}  \textfont\cmbsyfam=\eightcmbsy
  \scriptfont\cmbsyfam=\sixcmbsy \scriptscriptfont\cmbsyfam=\fivecmbsy
  \def\cmcsc{\fam\cmcscfam\eightcmcsc} \scriptfont\cmcscfam=\eightcmcsc
  \textfont\cmcscfam=\eightcmcsc \scriptscriptfont\cmcscfam=\eightcmcsc
     \fi             \tt \ttglue=.5em plus.25em minus.15em
  \normalbaselineskip=9pt
  \setbox\strutbox=\hbox{\vrule height7pt depth2pt width0pt}
  \let\sc=\sixrm \let\big=\eightbig \normalbaselines\rm }
\gdef\tenbig#1{{\hbox{$\left#1\vbox to8.5pt{}\right.\n@space$}}}
\gdef\ninebig#1{{\hbox{$\textfont0=\tenrm\textfont2=\tensy
   \left#1\vbox to7.25pt{}\right.\n@space$}}}
\gdef\eightbig#1{{\hbox{$\textfont0=\ninerm\textfont2=\ninesy
   \left#1\vbox to6.5pt{}\right.\n@space$}}}
\def\alternativefont#1#2{\ifx\arisposta\amsrisposta \relax \else
\xdef#1{#2} \fi}
\global\contaeuler=0 \global\contacyrill=0 \global\contaams=0
%
%
%
%
\newbox\fotlinebb \newbox\hedlinebb \newbox\leftcolumn
\gdef\makeheadline{\vbox to 0pt{\vskip-22.5pt
     \fullline{\vbox to8.5pt{}\the\headline}\vss}\nointerlineskip}
\gdef\makehedlinebb{\vbox to 0pt{\vskip-22.5pt
     \fullline{\vbox to8.5pt{}\copy\hedlinebb\hfil
     \line{\hfill\the\headline\hfill}}\vss} \nointerlineskip}
\gdef\makefootline{\baselineskip=24pt \fullline{\the\footline}}
\gdef\makefotlinebb{\baselineskip=24pt
    \fullline{\copy\fotlinebb\hfil\line{\hfill\the\footline\hfill}}}
\gdef\doubleformat{\shipout\vbox{\Landspec\makehedlinebb
     \fullline{\box\leftcolumn\hfil\columnbox}\makefotlinebb}
     \advancepageno}
\gdef\columnbox{\leftline{\pagebody}}
\gdef\line#1{\hbox to\hsize{\hskip\leftskip#1\hskip\rightskip}}
\gdef\fullline#1{\hbox to\fullhsize{\hskip\leftskip{#1}%
\hskip\rightskip}}
\gdef\footnote#1{\let\@sf=\empty
         \ifhmode\edef\#sf{\spacefactor=\the\spacefactor}\/\fi
         #1\@sf\vfootnote{#1}}
\gdef\vfootnote#1{\insert\footins\bgroup
         \ifnum\dimnota=1  \eightpoint\fi
         \ifnum\dimnota=2  \ninepoint\fi
         \ifnum\dimnota=0  \tenpoint\fi
         \interlinepenalty=\interfootnotelinepenalty
         \splittopskip=\ht\strutbox
         \splitmaxdepth=\dp\strutbox \floatingpenalty=20000
         \leftskip=\oldssposta \rightskip=\olddsposta
         \spaceskip=0pt \xspaceskip=0pt
         \ifnum\sinnota=0   \textindent{#1}\fi
         \ifnum\sinnota=1   \item{#1}\fi
         \footstrut\futurelet\next\fo@t}
\gdef\fo@t{\ifcat\bgroup\noexpand\next \let\next\f@@t
             \else\let\next\f@t\fi \next}
\gdef\f@@t{\bgroup\aftergroup\@foot\let\next}
\gdef\f@t#1{#1\@foot} \gdef\@foot{\strut\egroup}
\gdef\footstrut{\vbox to\splittopskip{}}
\skip\footins=\bigskipamount
\count\footins=1000  \dimen\footins=8in
\catcode`@=12
\tenpoint
\ifnum\unoduecol=1 \hsize=\tothsize   \fullhsize=\tothsize \fi
\ifnum\unoduecol=2 \hsize=\collhsize  \fullhsize=\tothsize \fi
\global\let\lrcol=L      \ifnum\unoduecol=1
\output{\plainoutput{\ifnum\tipbnota=2 \clearnmbnota\fi}} \fi
\ifnum\unoduecol=2 \output{\if L\lrcol
     \global\setbox\leftcolumn=\columnbox
     \global\setbox\fotlinebb=\line{\hfill\the\footline\hfill}
     \global\setbox\hedlinebb=\line{\hfill\the\headline\hfill}
     \advancepageno  \global\let\lrcol=R
     \else  \doubleformat \global\let\lrcol=L \fi
     \ifnum\outputpenalty>-20000 \else\dosupereject\fi
     \ifnum\tipbnota=2\clearnmbnota\fi }\fi
\def\ifdoublepage{\ifnum\unoduecol=2 }
\gdef\yespagenumbers{\footline={\hss\tenrm\folio\hss}}
\gdef\ciao{ \ifnum\fdefcontre=1 \endfdef\fi
     \par\vfill\supereject \ifnum\unoduecol=2
     \if R\lrcol  \headline={}\nopagenumbers\null\vfill\eject
     \fi\fi \end}

\newskip\olddsposta \newskip\oldssposta
\global\oldssposta=\leftskip \global\olddsposta=\rightskip

\def\filldots{\leaders\hbox to 1em{\hss.\hss}\hfill}
\def\inquadrb#1 {\vbox {\hrule  \hbox{\vrule \vbox {\vskip .2cm
    \hbox {\ #1\ } \vskip .2cm } \vrule  }  \hrule} }
 \def\newline{\hfil\break}
\def\jump{\vskip\baselineskip} \newskip\iinnffrr
\def\sjump{\iinnffrr=\baselineskip
          \divide\iinnffrr by 2 \vskip\iinnffrr}
\def\bjump{\vskip\baselineskip \vskip\baselineskip}
\newcount\nmbnota  \def\clearnmbnota{\global\nmbnota=0}
\newcount\tipbnota \def\letterfootnote{\global\tipbnota=1}

\def\note#1{\global\advance\nmbnota by 1 \ifnum\tipbnota=1
    \footnote{$^{\rm\nttlett}$}{#1} \else {\ifnum\tipbnota=2
    \footnote{$^{\nttsymb}$}{#1}
    \else\footnote{$^{\the\nmbnota}$}{#1}\fi}\fi}
\def\nttlett{\ifcase\nmbnota \or a\or b\or c\or d\or e\or f\or
g\or h\or i\or j\or k\or l\or m\or n\or o\or p\or q\or r\or
s\or t\or u\or v\or w\or y\or x\or z\fi}
\def\nttsymb{\ifcase\nmbnota \or\dag\or\sharp\or\ddag\or\star\or
\natural\or\flat\or\clubsuit\or\diamondsuit\or\heartsuit
\or\spadesuit\fi}   \clearnmbnota
\def\numberfootnote{\global\tipbnota=0} \numberfootnote
\def\setnote#1{\expandafter\xdef\csname#1\endcsname{
\ifnum\tipbnota=1 {\rm\nttlett} \else {\ifnum\tipbnota=2
{\nttsymb} \else \the\nmbnota\fi}\fi} }
\newcount\nbmfig  \def\clearnbmfig{\global\nbmfig=0}
\gdef\figure{\global\advance\nbmfig by 1
      {\rm fig. \the\nbmfig}}   \clearnbmfig
\def\setfig#1{\expandafter\xdef\csname#1\endcsname{fig. \the\nbmfig}}
 \def\endformula{\eqno\numero $$}
 \def\efr{\endformula}
\newcount\frmcount \def\clearfrmcount{\global\frmcount=0}
\def\numero{\global\advance\frmcount by 1   \ifnum\indappcount=0
  {\ifnum\cpcount <1 {\hbox{\rm (\the\frmcount )}}  \else
  {\hbox{\rm (\the\cpcount .\the\frmcount )}} \fi}  \else
  {\hbox{\rm (\applett .\the\frmcount )}} \fi}
\def\nameformula#1{\global\advance\frmcount by 1%
\ifnum\draftnum=0  {\ifnum\indappcount=0%
{\ifnum\cpcount<1\xdef\spzzttrra{(\the\frmcount )}%
\else\xdef\spzzttrra{(\the\cpcount .\the\frmcount )}\fi}%
\else\xdef\spzzttrra{(\applett .\the\frmcount )}\fi}%
\else\xdef\spzzttrra{(#1)}\fi%
\expandafter\xdef\csname#1\endcsname{\spzzttrra}
\eqno \hbox{\rm\spzzttrra} $$}
\def\nfr{\nameformula}    
\def\nameali#1{\global\advance\frmcount by 1%
\ifnum\draftnum=0  {\ifnum\indappcount=0%
{\ifnum\cpcount<1\xdef\spzzttrra{(\the\frmcount )}%
\else\xdef\spzzttrra{(\the\cpcount .\the\frmcount )}\fi}%
\else\xdef\spzzttrra{(\applett .\the\frmcount )}\fi}%
\else\xdef\spzzttrra{(#1)}\fi%
\expandafter\xdef\csname#1\endcsname{\spzzttrra}
  \hbox{\rm\spzzttrra} }      \clearfrmcount
\newcount\cpcount \def\clearcpcount{\global\cpcount=0}
\newcount\subcpcount \def\clearsubcpcount{\global\subcpcount=0}
\newcount\appcount \def\clearappcount{\global\appcount=0}
\newcount\indappcount \def\clearindappcount{\indappcount=0}
\newcount\sottoparcount 

\def\applett{\ifcase\appcount  \or {A}\or {B}\or {C}\or
{D}\or {E}\or {F}\or {G}\or {H}\or {I}\or {J}\or {K}\or {L}\or
{M}\or {N}\or {O}\or {P}\or {Q}\or {R}\or {S}\or {T}\or {U}\or
{V}\or {W}\or {X}\or {Y}\or {Z}\fi    \ifnum\appcount<0
\immediate\write16 {Panda ERROR - Appendix: counter "appcount"
out of range}\fi  \ifnum\appcount>26  \immediate\write16 {Panda
ERROR - Appendix: counter "appcount" out of range}\fi}
\clearappcount  \clearindappcount \newcount\connttrre
\def\clearconnttrre{\global\connttrre=0} \newcount\countref
\def\clearcountref{\global\countref=0} \clearcountref
\def\chapter#1{\global\advance\cpcount by 1 \clearfrmcount
                 \goodbreak\null\vbox{\jump\nobreak
                 \clearsubcpcount\clearindappcount
                 \itemitem{\ttaarr\the\cpcount .\qquad}{\ttaarr #1}
                 \par\nobreak\jump\sjump}\nobreak}
\def\section#1{\global\advance\subcpcount by 1 \goodbreak\null
               \vbox{\sjump\nobreak\ifnum\indappcount=0
                 {\ifnum\cpcount=0 {\itemitem{\ppaarr
               .\the\subcpcount\quad\enskip\ }{\ppaarr #1}\par} \else
                 {\itemitem{\ppaarr\the\cpcount .\the\subcpcount\quad
                  \enskip\ }{\ppaarr #1} \par}  \fi}
                \else{\itemitem{\ppaarr\applett .\the\subcpcount\quad
                 \enskip\ }{\ppaarr #1}\par}\fi\nobreak\jump}\nobreak}
\clearsubcpcount
\def\appendix#1{\global\advance\appcount by 1 \clearfrmcount
                  \goodbreak\null\vbox{\jump\nobreak
                  \global\advance\indappcount by 1 \clearsubcpcount
          \itemitem{ }{\hskip-40pt\ttaarr Appendix\ #1}
             \nobreak\jump\sjump}\nobreak}
\clearappcount \clearindappcount
\def\references{\goodbreak\null\vbox{\jump\nobreak
   \itemitem{}{\ttaarr References} \nobreak\jump\sjump}\nobreak}

\clearcpcount\clearcountref

\def\setchap#1{\ifnum\indappcount=0{\ifnum\subcpcount=0%
\xdef\spzzttrra{\the\cpcount}%
\else\xdef\spzzttrra{\the\cpcount .\the\subcpcount}\fi}
\else{\ifnum\subcpcount=0 \xdef\spzzttrra{\applett}%
\else\xdef\spzzttrra{\applett .\the\subcpcount}\fi}\fi
\expandafter\xdef\csname#1\endcsname{\spzzttrra}}
\newcount\draftnum \newcount\ppora   \newcount\ppminuti
\global\ppora=\time   \global\ppminuti=\time
\global\divide\ppora by 60  \draftnum=\ppora
\multiply\draftnum by 60    \global\advance\ppminuti by -\draftnum
\def\droggi{\number\day /\number\month /\number\year\ \the\ppora
:\the\ppminuti}     \global\draftnum=0
\def\draftcomment#1{\ifnum\draftnum=0 \relax \else
{\ {\bf ***}\ #1\ {\bf ***}\ }\fi} 
%
%
\catcode`@=11
\gdef\Ref#1{\expandafter\ifx\csname @rrxx@#1\endcsname\relax%
{\global\advance\countref by 1    \ifnum\countref>200
\immediate\write16 {Panda ERROR - Ref: maximum number of references
exceeded}  \expandafter\xdef\csname @rrxx@#1\endcsname{0}\else
\expandafter\xdef\csname @rrxx@#1\endcsname{\the\countref}\fi}\fi
\ifnum\draftnum=0 \csname @rrxx@#1\endcsname \else#1\fi}
\gdef\beginref{\ifnum\draftnum=0  \gdef\Rref{\fairef}
\gdef\endref{\scriviref} \else\relax\fi
\ifx\risposta\mplarisposta \ninepoint \fi
\parskip 2pt plus.2pt \baselineskip=12pt}
\def\Reflab#1{[#1]} \gdef\Rref#1#2{\item{\Reflab{#1}}{#2}}
\gdef\endref{\relax}  \newcount\conttemp
\gdef\fairef#1#2{\expandafter\ifx\csname @rrxx@#1\endcsname\relax
{\global\conttemp=0 \immediate\write16 {Panda ERROR - Ref: reference
[#1] undefined}} \else
{\global\conttemp=\csname @rrxx@#1\endcsname } \fi
\global\advance\conttemp by 50  \global\setbox\conttemp=\hbox{#2} }
\gdef\scriviref{\clearconnttrre\conttemp=50
\loop\ifnum\connttrre<\countref \advance\conttemp by 1
\advance\connttrre by 1
\item{\Reflab{\the\connttrre}}{\unhcopy\conttemp} \repeat}
\clearcountref \clearconnttrre
\catcode`@=12
\ifx\risposta\mplarisposta \def\Reflab#1{#1.} \letterfootnote \fi

\def\slashchar#1{\setbox0=\hbox{$#1$} \dimen0=\wd0
     \setbox1=\hbox{/} \dimen1=\wd1 \ifdim\dimen0>\dimen1
      \rlap{\hbox to \dimen0{\hfil/\hfil}} #1 \else
      \rlap{\hbox to \dimen1{\hfil$#1$\hfil}} / \fi}
\ifx\oldchi\undefined \let\oldchi=\chi
  \def\cchi{{\raise 1pt\hbox{$\oldchi$}}} \let\chi=\cchi \fi

\def\frac#1#2{{\textstyle{#1 \over #2}}}

\def\half{\ifinner {\scriptstyle {1 \over 2}}\else {1 \over 2} \fi}

\def\simge{\rlap{\raise 2pt \hbox{$>$}}{\lower 2pt \hbox{$\sim$}}}
\def\simle{\rlap{\raise 2pt \hbox{$<$}}{\lower 2pt \hbox{$\sim$}}}

\def\vbig#1#2{{\vbigd@men=#2\divide\vbigd@men by 2%
\hbox{$\left#1\vbox to \vbigd@men{}\right.\n@space$}}}

%
%
\newcount\fdefcontre \newcount\fdefcount \newcount\indcount
\newread\filefdef  \newread\fileftmp  \newwrite\filefdef
\newwrite\fileftmp     \def\strip#1*.A {#1}
\def\futuredef#1{\beginfdef
\expandafter\ifx\csname#1\endcsname\relax%
{\immediate\write\fileftmp {#1*.A}
\immediate\write16 {Panda Warning - fdef: macro "#1" on page
\the\pageno \space undefined}
\ifnum\draftnum=0 \expandafter\xdef\csname#1\endcsname{(?)}
\else \expandafter\xdef\csname#1\endcsname{(#1)} \fi
\global\advance\fdefcount by 1}\fi   \csname#1\endcsname}

\def\beginfdef{\ifnum\fdefcontre=0
\immediate\openin\filefdef \jobname.fdef
\immediate\openout\fileftmp \jobname.ftmp
\global\fdefcontre=1  \ifeof\filefdef \immediate\write16 {Panda
WARNING - fdef: file \jobname.fdef not found, run TeX again}
\else \immediate\read\filefdef to\spzzttrra
\global\advance\fdefcount by \spzzttrra
\indcount=0      \loop\ifnum\indcount<\fdefcount
\advance\indcount by 1   \immediate\read\filefdef to\spezttrra
\immediate\read\filefdef to\sppzttrra
\edef\spzzttrra{\expandafter\strip\spezttrra}
\immediate\write\fileftmp {\spzzttrra *.A}
\expandafter\xdef\csname\spzzttrra\endcsname{\sppzttrra}
\repeat \fi \immediate\closein\filefdef \fi}
\def\endfdef{\immediate\closeout\fileftmp   \ifnum\fdefcount>0
\immediate\openin\fileftmp \jobname.ftmp
\immediate\openout\filefdef \jobname.fdef
\immediate\write\filefdef {\the\fdefcount}   \indcount=0
\loop\ifnum\indcount<\fdefcount    \advance\indcount by 1
\immediate\read\fileftmp to\spezttrra
\edef\spzzttrra{\expandafter\strip\spezttrra}
\immediate\write\filefdef{\spzzttrra *.A}
\edef\spezttrra{\string{\csname\spzzttrra\endcsname\string}}
\iwritel\filefdef{\spezttrra}
\repeat  \immediate\closein\fileftmp \immediate\closeout\filefdef
\immediate\write16 {Panda Warning - fdef: Label(s) may have changed,
re-run TeX to get them right}\fi}
\def\iwritel#1#2{\newlinechar=-1
{\newlinechar=`\ \immediate\write#1{#2}}\newlinechar=-1}
\global\fdefcontre=0 \global\fdefcount=0 \global\indcount=0
%
%
\null
%
%
%
%

%
\loadamsmath
\loadeuler 
\mathchardef\bbalpha="710B
\mathchardef\bbbeta="710C
\mathchardef\bbgamma="710D
\mathchardef\bbomega="7121
\def\balpha{{\bfmath\bbalpha}}
\def\bbeta{{\bfmath\bbbeta}}
\def\bgamma{{\bfmath\bbgamma}}
\def\bomega{{\bfmath\bbomega}}
\def\U{{\rm U}}
\def\SU{{\rm SU}}
\pageno=0\baselineskip=14pt
\nopagenumbers{
\line{\hfill SWAT/96, CBPF-NF-001/96}
\line{\hfill\tt hep-th/9512116}
\line{\hfill December 1995}
\ifdoublepage \bjump\bjump\bjump\bjump\else\vfill\fi
\centerline{\capsone Non-abelian duality in 
$N=4$ supersymmetric gauge theories}
\bjump\sjump
\centerline{\scaps Nicholas Dorey, Christophe Fraser, Timothy J. Hollowood} 
\sjump
\centerline{\sl Department of Physics, University of Wales Swansea,}
\centerline{\sl Singleton Park, Swansea SA2 8PP, U.K.}
\centerline{\tt n.dorey, c.fraser, t.hollowood @swansea.ac.uk}
\sjump
\centerline{and}
\sjump
\centerline{\scaps Marco A. C. Kneipp}
\sjump
\centerline{\sl Centro Brasileiro de Pesquisas Fisicas (CBPF),
Rua Dr. Xavier Sigand, 150}
\centerline{\sl 22290 Rio de Janeiro, Brazil}
\centerline{\tt kneipp@cbpfsu1.cat.cbpf.br}
\bjump\bjump
\ifdoublepage
\vfill
\eject\null\vfill\fi
\centerline{\capsone ABSTRACT}\sjump
A semi-classical check of the Goddard-Nuyts-Olive (GNO) generalized duality
conjecture for gauge theories with adjoint Higgs fields is performed 
for the case where the unbroken gauge group is non-abelian. The
monopole solutions of the theory transform under the
non-abelian part of the unbroken global symmetry and the associated
component of the 
moduli space is a Lie group coset space. The well-known problems
in introducing collective coordinates for these degrees-of-freedom are
solved by considering suitable multi-monopole configurations 
in which the long-range 
non-abelian fields cancel. In the context of an $N=4$ supersymmetric gauge
theory, the multiplicity of BPS saturated states is given by the number of
ground-states of a supersymmetric quantum mechanics on the compact
internal moduli space. The resulting degeneracy is expressed as the 
Euler character of the coset space. In all cases the
number of states is consistent with the dimensions of the 
multiplets of the unbroken dual gauge group, and hence the results
provide strong support for the GNO conjecture. 
\sjump\vfill
\eject}
\yespagenumbers\pageno=1
%
%

\chapter{Introduction}

There is now strong evidence [\Ref{SEN},\Ref{EV}] 
that a version of the electromagnetic
duality originally conjectured by Montonen and Olive [\Ref{MO}] is exactly
realized in $N=4$ supersymmetric gauge theory.\note{The relevance of
$N=4$ supersymmetry in this context was first suggested by 
Osborn [\Ref{OS}].} In its simplest form, 
duality requires that the spectrum of BPS saturated states 
in the theory should be invariant under the interchange of
electric and magnetic quantum numbers $m \leftrightarrow n$ (here we use
the notation of [\Ref{SEN}]) together with the inversion of
the gauge coupling constant $e \rightarrow 2\pi/e$. 
This means that a state with quantum numbers
$(m,n)$ must have a distinct partner with quantum numbers
$(n,m)$ whenever $m\neq n$. In principle, 
this condition can be checked reliably at weak coupling 
using standard semi-classical methods for the quantization of
monopoles. Sen provided an important confirmation of 
duality in the case of gauge group $\SU(2)$ broken to U(1) 
when he demonstrated the
existence of a bound-state in the two-monopole sector with quantum
numbers $(1,2)$ which is the partner of a particular Julia-Zee 
dyon.\note{Actually Sen showed the existence of a whole tower of states
with quantum numbers $(n,2)$ where $n$ is any odd integer whose
existence is required by the full ${\rm SL}(2,{\Bbb Z})$ duality. In this
paper we will set $\theta=0$ and discuss only the ${\Bbb Z}_{2}$ subgroup
originally considered by Montonen and Olive.}

The Montonen-Olive duality conjecture for gauge group $\SU(2)$ was
generalized to the case of an arbitrary compact Lie group by Goddard,
Nuyts and Olive (GNO) in [\Ref{GNO}]. 
The GNO conjecture states that a gauge theory
with gauge group $G$, has a dual description at strong coupling
in terms of a weakly coupled gauge theory with gauge group
$G^{\vee}$. The dual gauge group is defined by requiring that its 
Lie algebra $g^{\vee}$ has roots which are the duals of the root of
$g$ defined by $\balpha^\vee=2\balpha/\balpha^2$.
More precisely, we should also specify that $G$ is
spontaneously broken to a subgroup $H$ by a Higgs scalar in the
adjoint representation so that the theory contains magnetic 
monopoles. GNO duality relates the spectrum of BPS saturated
monopole states in this theory to the spectrum of massive gauge bosons
in the dual gauge theory where $G^{\vee}$ breaks to $H^{\vee}$. A
natural question to ask is whether this generalized electromagnetic duality
is an exact relation between the corresponding 
$N=4$ supersymmetric gauge theories. Assuming the same states are
present for all values of the coupling, such a relation places a
powerful constraint on the semi-classical spectrum of BPS monopoles. 
This is particularly apparent when the unbroken subgroups $H$
and $H^{\vee}$ contain  
non-abelian factors. 
In this case the gauge bosons form multiplets of the unbroken
non-abelian symmetry leading to degeneracies in the particle spectrum.
In order for GNO duality to hold it is necessary that these
degeneracies are precisely matched in the semi-classical spectrum of
BPS monopoles in the dual theory. In this paper 
we will argue that this constraint is indeed satisfied. 

Our aim is to determine the semi-classical spectrum of magnetic
monopoles in a supersymmetric 
gauge theory with an unbroken non-abelian subgroup. 
However there are some well-known theoretical obstacles to applying
the standard semi-classical reasoning in this case which we now review.      
In general a particular monopole solution breaks several of the
global symmetries of the theory. The action of these symmetries
usually gives rise to normalizable zero-modes of the Hamiltonian for small
fluctuations around the monopole
background. The zero-modes can be eliminated from the path integral 
by introducing collective
coordinates which parametrize a moduli space of gauge-inequivalent
solutions. A familiar example is the translational and
charge rotation degrees-of-freedom of the 't Hooft-Polyakov 
monopole which form a moduli space isometric to 
${\Bbb R}^{3}\times S^{1}$. Semi-classical quantization of these
degrees-of-freedom, yields a tower of massive particle states with integer
electric charges. When the unbroken gauge group is non-abelian, it
appears that  the monopole can also transform under some
generators of the corresponding non-abelian global symmetry. Na\"\i vely
one might expect that the resulting semi-classical spectrum would consist of 
multiplets of this symmetry (`chromo-dyons') 
analogous to the isospin multiplets which
contain the nucleon and $\Delta$ states in the Skyrme model.   
However the action of these symmetries does {\it not\/} give rise to any
additional normalizable zero-modes and collective coordinates cannot be
introduced in the usual way. 

The problem mentioned above has its origin in the slow
fall-off of the non-abelian field of the monopole at large
distance. In unitary gauge, the gauge field falls off like  $P/r$,
where $P$ is a constant element of the Lie algebra $h$ of the unbroken
subgroup $H$. The action of the symmetry generators $Q \in h$ for
which $[Q,P] \neq 0$ therefore produces a variation of the field which
also falls off like $1/r$. At first sight the problem seems simple; 
the zero-modes corresponding to such a variation would have linearly
divergent norm. Assuming this to be the case, the moment-of-inertia
for motion in the symmetry directions would be infinite and the
semi-classical spectrum would contain an infinite number of degenerate
chromo-dyons.\note{A similar situation is known to occur for solitons of the
${\rm O}(3)$ non-linear $\sigma$-model in (2+1)-dimensions
[\Ref{WL}].} However,
Abouelsaoud showed that this scenario is incorrect
[\Ref{A3},\Ref{A2}]. In a gauge
theory, the zero-modes must not only obey the linearized field equation
but should also preserve the local gauge condition. For a zero-mode
which corresponds to the non-trivial 
action of a global non-abelian symmetry generator on the
large-distance field of the monopole, these two conditions cannot be
satisfied simultaneously. In fact, in the case at hand, 
the magnetic monopole 
{\it does\/} have extra non-normalizable zero-modes which satisfy the 
gauge condition. However these modes fall off much slower
than $1/r$ at large distance, and so they cannot be associated
to the action of the non-abelian symmetry generators. Weinberg
suggested that these modes should be thought of instead as the bottom of 
the continuum of scattering eigenstates [\Ref{WB1}]. The occurrence of these 
pathologies reflects a deeper problem with the very notion of global 
non-abelian symmetry in the monopole sector which was uncovered by
Nelson and Manohar [\Ref{NM}] (see also [\Ref{N}]) 
and by Balachandran et al [\Ref{BMMNS1}]. 
These authors proved that it is impossible to find a 
set of generators which represent the unbroken global subgroup and are
well-defined at every point on a large sphere containing the
monopole. The obstruction to finding such a set is topological in
nature and the result is somewhat analogous to the `hairy ball' 
theorem which states that there are no smooth, non-vanishing 
vector fields on the $2$-sphere.   

In the light of the above, it is not clear how to apply the
semi-classical method to find the spectrum of states associated with a
single monopole in isolation. A simple way to bypass these problems,
which was suggested by Coleman and Nelson [\Ref{NC}], is to consider instead a
chromo-magnetic dipole; a non-abelian monopole ($M$) and anti-monopole
($\bar{M}$), whose long-range gauge fields exactly cancel.\note{In
the following, the term `long-range' always indicates the term of
order $1/r$ in the large-distance expansion of the gauge and Higgs
fields.} If the separation, $R$, between $M$ and $\bar{M}$ is much 
greater than the core-size, this configuration is well approximated by a linear
superposition of the corresponding gauge fields. In the absence of
long-range fields, there is no obstacle to defining the action of the
global symmetry generators on the field configuration.      
Importantly, Coleman and Nelson prove
that the corresponding zero-modes now have finite norm, proportional to
$R$. These modes can now be eliminated by introducing 
collective coordinates in the usual way 
and the resulting moduli space is a coset 
space. The finite norms of the zero-modes means that the 
moments-of-inertia for motion on this space are also finite. The
semi-classical spectrum that arises from quantizing this motion
includes a tower of chromo-dyon states which carry non-abelian
charge. However these states are not
localized around either the monopole or the anti-monopole, rather they
are to be thought of as excitations of the colour flux tube joining
the two sources, and they become
increasingly hard to distinguish from the ground-state as $R\rightarrow
\infty$. 
Unfortunately in the BPS limit the configuration of Coleman and Nelson
is not appropriate because the Higgs field now has a long-range
component in the $M-\bar M$ system which cannot be cancelled.

In the following we will pursue a simple variant of Coleman and
Nelson's idea. As the problem lies in the component of the long-range field
which fails to commute with the generators of $h$, we must certainly
consider a configuration in which this is cancelled at order $1/r$. 
However, if one does not also require that the other long-range 
components of the field cancel, then this condition can be satisfied
by considering a configuration which contains {\it only} monopoles and no
anti-monopoles. Such a configuration has the advantage that it 
saturates the Bogomol'nyi bound and therefore corresponds to an exact
solution of the field equations. However, when the unbroken gauge
group has complex representations, there is a 
price which one must pay for this simplification. For example, 
in the case where
the unbroken gauge group contains a factor $K=\SU(N)$, 
there are ${\rm rank}[K]=N-1$ independent non-abelian components or 
hypercharges which must be cancelled and one is forced to consider 
a configuration of $N$ BPS monopoles. Despite this complication we
will be able to obtain the semi-classical spectrum of the system by
arguments very similar to those used by Coleman and Nelson and
generalize the analysis to the supersymmetric case. 

The aim of our analysis is to provide a semi-classical test of GNO
duality. With this in mind, it is useful to consider two slightly different 
physical interpretations of the semiclassical spectrum associated with
the multi-monopole configuration described above. The first point of
view, which is the closer in spirit to [\Ref{NC}], is to focus on a
single monopole and to 
regard the additional monopoles, which are necessary to cancel the
long-range non-abelian fields, as a regulator which can be removed to
infinity at the end of the calculation. From this point of view, 
it is convenient (but not necessary) to think of the multi-monopole
solution as a single monopole separated by a distance $R$ 
from a conglomerate consisting 
of all the other monopoles sitting at the same point. The geometry of
such a configuration is then identical to that of the monopole and
anti-monopole of Coleman and Nelson. 
The global colour transformations which affect the first 
monopole lead to a moduli space which has the form of a coset.
Our aim is to quantize the collective coordinate motion on this space
and deduce the spectrum of quantum states of the monopole in the
presence of the monopole conglomerate which cancels the long-range non-abelian 
fields.\note{In this paper 
we will explicitly consider only 
electrically neutral states of the individual monopoles and postpone
the analysis of electric dyon states for future work.} 
From the degeneracy of these states in the limit
$R\rightarrow \infty$ we can then infer the 
degeneracy of BPS saturated states of a single monopole which, 
according to the GNO conjecture, must 
match those of the gauge bosons in the dual theory. 

Alternatively, one may take the point of view that 
the problems discussed above for states with long-range non-abelian fields
are fundamental in nature and we should only consider the
sector of the Hilbert space in which these fields are absent. 
This means restricting our attention to 
multi-particle states which carry 
zero total hypercharge both with respect to the unbroken non-abelian
gauge group $K$ and its dual $K^{\vee}$.  
By acting on the multi-monopole configuration described above 
with the global symmetry generators we can construct a
moduli space of solutions whose long-range magnetic fields are purely
abelian. Quantizing motion on this moduli space we will find a
spectrum of BPS-saturated states, consisting of $N$ fundamental monopoles, 
which carry zero hypercharge with respect to the unbroken
chromo-magnetic gauge group $K^{\vee}$. 
These states are dual to states, containing the same number of gauge
bosons which carry zero total hypercharge with respect to the unbroken 
chromo-electric gauge group, $K$. We should stress that the states we
are considering comprise of monopoles or of gauge bosons at arbitrarily
large separations: they are multi-particle states and have nothing to
do with the stable bound states of the multi-monopole system discussed
in [\Ref{SEN}]. Hence our proposed test of GNO
duality can be interpreted in two slightly different ways. The first
approach, in which the additional monopoles are regarded as a
regulator, allows us to define the semiclassical spectrum associated
with a single monopole while in the second we formulate duality
directly between composite states with zero non-abelian hypercharge.  
In fact the degeneracies required for GNO duality can
be demonstrated equally well in either approach. However, for
simplicity of presentation, we will concentrate on the former
approach in cases where more than two monopoles are required.       

In this expanded introductory section 
we sketch these ideas in the simplest
case; a theory with gauge group $G=\SU(3)$ which is broken down to
$H=\SU(2)\times \U(1)$ by an adjoint Higgs mechanism. In the remainder of
the paper, we consider the general case of 
an arbitrary compact gauge group with
arbitrary pattern of symmetry breaking. 
The full details of our conventions for the general case are given
in Section 2 below. However, to make this section as self-contained as
possible, we will introduce the relevant notation for the minimal case below.
 
We consider an $\SU(3)$ gauge group generated by Cartan elements 
$H_{i}$, $i=1,2$, and step
generators $E_{\pm\balpha}$, $E_{\pm\bbeta}$ and $E_{\pm\bgamma}$. 
The roots $\balpha$, $\bbeta$ and $\bgamma=\balpha+\bbeta$  satisfy 
$\balpha^{2}=\bbeta^{2}=\bgamma^{2}=2$ with $\balpha\cdot\bbeta
=-1$. An unbroken $\SU(2)\times \U(1)$ generated by
$E_{\pm\bbeta}$ together with the Cartan elements is selected by
chosing a vacuum expectation value for the Higgs, $\phi_{0}=i{\bfmath 
v}\cdot{\bfmath H}$ where ${\bfmath v}\cdot\bbeta=0$. Spherically
symmetric BPS monopole solutions of this theory were classified by Weinberg 
[\Ref{WB1},\Ref{WB2}] and are
obtained by embedding the standard $\SU(2)$ monopole into the gauge
group $\SU(3)$. 
Weinberg identified a set of fundamental monopole solutions which
cannot be decomposed into constituents of lower mass. The other
spherically symmetric 
solutions consist of a number of fundamental monopoles all sitting at
the same point in space.
Weinberg performed an analysis of the
spectrum of small fluctuations around these configurations and showed
that the number of zero frequency modes was consistent with precisely
this interpretation. In the following we will assume that this picture
is correct and, more generally, that there exist exact solutions of the
Bogomol'nyi equation describing fundamental monopoles at arbitrary
separations. For an arbitrary multi-monopole 
solution the long-range behaviour
of the fields in unitary gauge is just,
$$\eqalign{
B_i&={x_i\over r^3}{\cal G}+{\cal O}(1/r^{3}),\cr
\phi&=i{\bfmath v}\cdot{\bfmath H}-{{\cal G}\over r}+{\cal O}(1/r^{2}).\cr}
\nfr{ASY}  
By a global gauge transformation the non-abelian charge ${\cal G}$ can
be chosen to lie in the Cartan subalgebra, ${\cal G}=i{\bfmath
g}\cdot{\bfmath H}$. As we review in Section 4, 
the vector ${\bfmath g}$ is then constrained to lie in an integer
lattice by the Dirac quantization condition. 
In the current case this implies that, 
$$
{\bfmath g}= {1\over2e} \left(n_{\balpha}\balpha+n_{\bbeta}\bbeta \right)
\nfr{su2quant}
where $n_{\balpha}$ and $n_{\bbeta}$ are non-negative integers which we
will call magnetic weights. 

In our case, the fundamental monopoles have magnetic
weights $\{ n_{\balpha},n_{\bbeta}\}$ given by $\{1,0 \}$ and $\{1,1 \}$. These
configurations are degenerate in mass and are related by a global 
gauge transformation which lies in the Weyl subgroup of the unbroken
$\SU(2)$. In general each of these monopoles will have a long-range
field which fails to commute with the step generators
$E_{\pm\bbeta}$. However its easy to see that if we consider a field
configuration which consists of two well-separated 
monopoles, one of each type, 
the resulting long-range fields $\propto(2\balpha
+\bbeta)\cdot {\bfmath H} /r$ are now invariant under all the
generators of the unbroken subgroup $H$. Dipole terms which do not
commute with $E_{\pm\bbeta}$ arise at ${\cal O}(1/r^{2})$ and the
action of these generators would now appear to give zero-modes which have the
same fall off and are therefore normalizable. In the case of the
monopole-anti-monopole, system Coleman and Nelson [\Ref{NC}] were able to prove
that this na\"\i ve conclusion is in fact correct. In Section 6 we will
briefly review their arguments and claim that the proof given in [\Ref{NC}]
should apply equally to our multi-monopole configurations. For the
moment we will assume that this is the case. 

The action of the generators 
$\{H_{1},H_{2},E_{\bbeta}, E_{-\bbeta}\}$ 
of the unbroken global subgroup on the
two-monopole configuration naturally gives rise to a continuous
manifold of solutions. The analysis for the case at hand is simple; 
each monopole transforms under ordinary electric charge rotations
which are generated by $i{\bfmath v}\cdot{\bfmath H}$, however it is easy
to show that each 
monopole is invariant under the action of a linearly independent 
Cartan generator. This means that the only action of the Cartan
subalgebra is to generate independent charge rotations of each of the
two monopoles. Quantizing these degrees-of-freedom 
gives rise to a tower of electrically charged states or dyons
associated with either monopole. For the simplest check of GNO
duality, we only need to count the number of electrically neutral
BPS saturated states of the system and, for this reason we will ignore these
directions in the moduli space. Similarly, the six coordinates which
describe the centre-of-mass positions of the two monopoles can be
thought of as fixed for our purposes. Our manifold of solutions then is 
generated by the action of the two step generators $E_{\pm\bbeta}$ and
can therefore be identified with the coset space
${\cal M}=\SU(2)/\U(1)$.\note{In fact there is also a discrete
subgroup which must be divided out [\Ref{NC}]. However, as all the
states which we consider are $\SU(2)$ singlets, this will play
no role in our discussion.}  

The occurrence of a moduli space of dimension two seems to
present something of a puzzle. The linearized Bogomol'nyi equation has
a well known symmetry under the action of the three inequivalent
almost complex structures on four dimensional Euclidean 
space.\note{As we will review in
Section 7, the monopole can be thought of as configuration in 
a four dimensional Euclidean space where the bosonic fields form a
four-vector $W_{m}=(A_i, \phi)$.} (See for example [\Ref{G1}]). 
Under the action of this
discrete symmetry, zero-modes naturally come in multiplets of four. 
Familiar examples 
are the three translational and single charge rotational mode
of the monopole which naturally form such a set. In the $\SU(2)$ case, 
the almost complex
structures on ${\Bbb R}^{4}$ descend to give three inequivalent almost complex
structures on the moduli space which can then be shown to be a 
hyper-K\"{a}hler manifold. We shall argue that the
two zero-modes generated by the action of $E_{\pm\bbeta}$ 
are related to each other by a single complex structure
on ${\Bbb R}^{4}$ which endows the coset space $\SU(2)/\U(1)$ with its
canonical K\"{a}hler form. We will find that the other two independent
complex structures on ${\Bbb R}^{4}$ generate two further normalizable
zero-modes of the two monopole system. At the level of a linearized
analysis, it is not clear whether or not
new collective coordinates are required for these modes. The only
reasonable candidates for such coordinates 
would be the relative non-abelian group orientation between the two
monopoles. However these directions in function space would clearly
take us outside the subspace of configurations for which the
long-range non-abelian fields cancel, which appears to contradict the
fact that the corresponding modes are normalizable. In any
case, these modes do not correspond to the action of any unbroken global 
symmetry generator on the two-monopole system and we will
assume that their presence does not affect our conclusions.\note{See
Section 9 for further discussion of this issue.}
 
As usual the semi-classical spectrum is obtained by allowing the
collective coordinates which live on ${\cal M}=\SU(2)/\U(1)$ to become time
dependent [\Ref{MAN}]. 
As discussed above, variations of these coordinates give rise to
normalizable zero-modes of the configuration 
and the corresponding moments-of-inertia 
define a Riemannian metric on ${\cal M}$. For a coset space $K/L$, 
the metric is determined, up to a scale factor, by invariance under
natural action of $K$. The relevant scale factor is given by the norm
of the zero-modes which grows linearly with the separation $R$ between the
two monopoles. Quantizing this system is straightforward, the
resulting (electrically neutral) states transform in representations
of $\SU(2)$ labelled by integral spin $j$. The mass spectrum is 
$E=M_{\rm cl}+j(j+1)/2I$ where $M_{\rm cl}$ is
the classical mass of the two monopole system and $I\sim R/e^2$ is the 
moment-of-inertia. The
states with non-zero $j$ are analogous to the chromo-dyons identified 
by Coleman and Nelson. Because of the non-zero rotational contribution
to their energy, these states do not saturate the Bogomol'nyi bound
and hence they are not relevant to our proposed
test of duality. The main result of this analysis therefore is that
the two monopole system in this model 
has a unique BPS-saturated ground-state. In Section 4,  we will argue that
this result holds for an arbitrary gauge group and therefore the 
degeneracies in the semi-classical spectrum required for GNO duality are
not present in ordinary (non-supersymmetric) gauge theory. Of course 
supersymmetry is certainly
necessary anyway for exact GNO duality for many other reasons such as
the question of monopole spin [\Ref{OS}]. 
In the following we will show that it
also provides the necessary degeneracies of states to fill out the
appropriate multiplets of the dual gauge group.  

In a gauge theory with extended supersymmetry, BPS monopoles 
are coupled to massless Dirac fermions. 
The Callias index theorem [\Ref{CAL}] 
states that the fermion fields will have 
exact zero-modes in the monopole background. The number of these modes
is determined by the topological charge, as well as 
the number of massless fermions in the theory and
their transformation properties under the $\SU(2)$ subgroup in which
the monopole is embedded. Like their bosonic counterparts, these modes
can be eliminated from the path integral by introducing 
collective coordinates. For fermion zero-modes these coordinates are
time-dependent Grassman numbers and the resulting
semi-classical description now involves the quantum mechanics of these
degrees-of-freedom coupled to the bosonic coordinates on ${\cal M}$. 

In the case of extended supersymmetry, BPS monopole configurations 
have the important property that they are invariant under the action
of half the supersymmetry generators. These
unbroken supersymmetries provide a natural pairing between bosonic and
fermionic zero-modes. In the case of $\SU(2)$ gauge theory with $N=2$
supersymmetry, Gauntlett [\Ref{G1}] demonstrated that 
the unbroken supersymmetries in spacetime correspond directly to
unbroken supersymmetries in the collective coordinate 
quantum mechanics on the moduli space. These results were generalized
to the $N=4$ case by Blum [\Ref{BL}]. We will see below that a
similar phenomenon occurs for the semi-classical quantization of the 
global non-abelian degrees-of-freedom of the monopole described above. 

The compact moduli spaces 
${\cal M}$ which arise from the action of the global symmetry
generators in the general case are always a product of Lie group
coset spaces. Each of these factors has a natural 
complex structure which descends from a particular complex structure
on ${\Bbb R}^{4}$ which we discuss in Section 7. 
This complex structure is covariantly constant and
therefore each factor can be thought of independently 
as a K\"{a}hler manifold. Quantum mechanics on a K\"{a}hler manifold
can be endowed with up to four one-component supersymmetries 
($N=4\times {1\over 2}$) and our main result is that all four of
these worldline supersymmetries are realized precisely when the
overlying gauge theory has $N=4$ spacetime supersymmetry. These
supersymmetries determine the exact form of the collective coordinate
Lagrangian. The Hilbert space of supersymmetric ground-states of this system
corresponds to the de Rham complex of harmonic forms on ${\cal M}$ [\Ref{AG}]. 
As these ground-states necessarily have
zero energy, they correspond to inequivalent, electrically neutral 
BPS-saturated states of
the multi-monopole system. Importantly, because 
all the zero-modes associated with the action of the unbroken non-abelian 
gauge group, transform in the fundamental representation 
of the embedding $\SU(2)$, the states we consider all carry zero
angular momentum.\note{More precisely, each of these ground-states will
separately give rise to a small representation of supersymmetry after
the space-time degrees-of-freedom of the monopole are quantized.}

It turns out that in all cases of interest, the index of the de Rham complex 
is saturated by forms of even degree. This means that all the
resulting ground-states states are bosonic. Hence, the multiplicity of
these ground-states is simply equal to the Euler character, $\chi({\cal M})$,
of the internal moduli space. Using this result we can now 
check the GNO conjecture in the minimal
case described above. As the gauge group is simply-laced, the theory
should be self-dual. The spectrum of massive gauge bosons in this
model consists of an $\SU(2)$ doublet with 
electric charge $q=e{\balpha}\cdot\hat{\bfmath v}$ and
their anti-particles. Clearly the number of BPS-saturated states 
of electric charge $2q$ which carry zero $\SU(2)$ hypercharge is just
two. Thus for exact duality
we should find a two-fold degeneracy in the semi-classical spectrum of
our two monopole system at large separation. 
The relevant moduli space is just the
coset $\SU(2)/\U(1)$ which is homeomorphic to $S^{2}$. As
$\chi(S^{2})=2$, the result of our calculation is consistent with
duality. The wavefunction of the additional ground-state, which is not
present in the bosonic theory, is just the usual 
volume form on the coset space.  

In the remainder of the paper we show that a similar agreement holds in the
case of an arbitrary compact Lie group with arbitrary symmetry
breaking. The only possible exceptions to this agreement are the
so-called degenerate cases (see [\Ref{WB2},\Ref{WB4}]) 
where additional collective coordinates are known to arise. 
However, even in these cases, a na\"\i ve 
application of our results produces a spectrum which is consistent
with duality. The results provide
compelling evidence for GNO duality in $N=4$ supersymmetric gauge
theory. The paper is organized as follows: in Section 2 we discuss the
possible patterns of symmetry breaking in the general case. Section 3
is devoted to the spectrum of massive gauge bosons while Section 4
reviews the construction and classification of BPS monopoles in these
theories. In Section 5 we review the pathological situation for a
single monopole in isolation and in Section 6 we discuss how this can
be solved by considering the multi-monopole configuration described
above. Section 7 contains the discussion of supersymmetry and fermion
zero-modes. Finally in Section 8 we illustrate the consistency 
the semi-classical spectrum of monopoles with the expectations of the
GNO conjecture in a range of examples.    
  
\chapter{Symmetry breaking patterns}

In this section we consider the various symmetry breaking patterns
that are possible in a gauge theory with a single Higgs field in the adjoint
representation of the gauge group. For this purpose, we need only
consider a bosonic gauge theory,
with a compact simple Lie group $G$, and a single Higgs scalar field in
the adjoint representation. We choose a unitary gauge in which the Higgs field
on a large sphere at infinity is a constant $\phi_0$.
We can always choose 
a Cartan subalgebra of the Lie algebra $g$ of $G$ so that\note{Notice
that if the unbroken gauge group $H$ is
non-abelian then the choice of the Cartan subalgebra $\bfmath H$ is
not unique.}
$$
\phi_0=i{\bfmath v}\cdot{\bfmath H},
\nfr{HF}
where $\bfmath H$ are the Cartan elements of $g$ 
considered as an $r={\rm rank}(g)$ vector.
The Higgs field breaks the symmetry to a subgroup $H\subset G$
which consists of group elements which commute with the Higgs vacuum
expectation value:
$$
H=\left\{U\in G\vert\ U\phi_0U^{-1}=\phi_0\right\}.
\nfr{UGG}
The Lie
algebra $h$ of $H$ consists of the generators of $g$ commuting with
$\phi_0$. Introducing the usual Cartan-Weyl basis for the
complexification of $g$, consisting of Cartan elements $\bfmath H$ and
step generators $E_\balpha$, where $\balpha$ is a root of $g$,
then the elements of the complexification of $h$ are the Cartan
elements $\bfmath H$ and the step generators $E_\balpha$ with
$\balpha\cdot{\bfmath v}=0$.\note{Our
Lie algebra conventions are $[{\bfmath H},E_\balpha]=\balpha E_\balpha$,
$[E_\balpha,E_{-\balpha}]=2\balpha\cdot{\bfmath H}/\balpha^2$ and
${\bfmath H}^\dagger={\bfmath H}$ and $E_\balpha^\dagger=E_{-\balpha}$.
The elements of the real form of the 
Lie algebra will be chosen to be anti-hermitian.} Generically,
the unbroken gauge group will be the maximal torus U$(1)^r\subset
G$; however, if $\bfmath v$  
is orthogonal to any root of $g$ (i.e. $\phi_0$ is not {\it
regular\/}) the unbroken
gauge group will be non-abelian. 

Let $\balpha_i$, for $i=1,\ldots,r$, be
a set of simple roots of $g$, chosen so that
$\balpha_i\cdot{\bfmath v}\geq0$.  We can visualize the symmetry
breaking by means of the Dynkin diagram of $g$. The vector $\bfmath v$ can be
expanded in terms of the fundamental weights\note{These are
the $r$ vectors ${\bomega}_i$ such that
$\bomega_i\cdot\balpha _j=(\balpha_j^2/2)\delta_{ij}$.} 
$$
{\bfmath v}=\sum_{i=1}^r v_i{\bomega}_i.
\efr
The unbroken gauge group has the general form 
$$
H={{\rm U}(1)^{r'}\times K\over Z}
\efr
where $K$ is
semi-simple and the Dynkin diagram of $K$ is obtained from that of $g$ by
`knocking-out' spots corresponding to each non-zero $v_i$. So the
simple roots of $k$, the Lie algebra of $K$, are the subset of the
simple roots of $g$ orthogonal to $\bfmath v$.
$Z$ is a finite group which specifies the global structure of
$H$ and is isomorphic to a subgroup of the centre of $K$ [\Ref{GNO}].

\chapter{The spectrum of gauge bosons}

The classical spectrum of gauge bosons in such a model can
be calculated by diagonalizing the mass term induced by the Higgs
vacuum expectation value in the Lagrangian [\Ref{BAIS}]. This term is
$$
-{\rm Tr}\left([\phi_0,A_\mu]^2\right).
\efr
Associating the states with elements of the Lie algebra $g$
the states corresponding to the Cartan elements are massless while the
states associated to the step generators $E_\balpha$ have a mass
$$
M_\balpha=e\vert{\bfmath v}\cdot\balpha \vert,
\efr
which saturates the Bogomol'nyi bound since these states carry electric
charge $Q_e=e\hat{\bfmath v}\cdot\balpha$, where $\hat{\bfmath
v}={\bfmath v}/\vert{\bfmath v}\vert$.
Notice that the states corresponding to a root and its negative are degenerate.

The massless gauge bosons correspond to the unbroken gauge group $H$ and
the massive gauge bosons form multiplets of 
$H$. However, we expect that in the quantum theory some of the massive
states are unstable; namely the ones
for which $M^\pm_\bgamma \geq
M^\pm_\balpha +M^\pm_\bbeta $, with $\balpha $,
$\bbeta $ and $\bgamma $ all having non-zero inner
product with $\bfmath v$, and for which
there exists a coupling in the Lagrangian which can mediate the decay. 
Such a coupling exists whenever
$\bgamma =\balpha +\bbeta $, and so the
state associated to $\bgamma $ is at the threshold for decay if $\balpha$
and $\bbeta $ are either {\it both\/} positive roots or {\it both\/} 
negative roots.

The stable massive gauge bosons can be associated to sets of roots
of $G$, containing a simple root $\balpha_i$ with a non-zero inner
product with $\bfmath v$, which form representations of $H$
and whose lowest weight is
the simple root $\balpha_i$. The negative simple root $-\balpha_i$ 
is associated in
the same way with the complex conjugate representation.
In this way we associate stable massive gauge multiplets
with spots on the Dynkin diagram of $g$ which have been
`knocked-out'.\note{It has very recently been shown
[\Ref{G2},\Ref{W6}], that the gauge bosons at threshold do actually form 
bound-states which match bound-states of monopoles in the dual 
picture---see the Addendum. This fact does not directly affect our
conclusions.}

\chapter{Classical monopole solutions}

In this section we explain how BPS monopole solutions can be constructed
for the theories that we are considering and how the the
Dirac quantization condition is applied.

It is convenient to work in a unitary
gauge in which the Higgs field field is constant on the sphere at
infinity. Such a gauge can be achieved in a non-singular way by
working in the two-patch formalism of [\Ref{WY}]. In this gauge
there is a canonical definition of the unbroken
gauge group $H$, unlike in the radial gauge of [\Ref{TP}], where
the unbroken gauge group lies
inside $G$ in a position dependent way.

For a multi-monopole configuration the asymptotic form of the magnetic
field $B_i$ and the Higgs field, in this gauge are
$$\eqalign{
B_i&={x_i\over r^3}{\cal G}+{\cal O}\left({1\over r^{3}}\right),\cr
\phi&=i{\bfmath v}\cdot{\bfmath H}-{{\cal G}\over r}+{\cal O}\left({1
\over r^{2}} \right),\cr}
\nfr{ASY}
where ${\cal G}$ is a constant element of the Lie algebra $h$ and so
commutes with the asymptotic value of the Higgs field. 
Notice that in the BPS
limit the Higgs field has an ${\cal O}(1/r)$ component.
We can use the freedom present in the definition of the Cartan
subalgebra of $g$ (see section 2) to have 
$$
{\cal G}=i{\bfmath g}\cdot{\bfmath H},
\efr
for a real vector $\bfmath g$. The Dirac quantization can now be stated in
a simple way [\Ref{GNO},\Ref{EW}]. The vector $\bfmath g$, up to a normalizing
factor, must lie in the co-root lattice of $g$ denoted $\Lambda_R^\vee$.
This is the lattice defined as the integer span of the simple co-roots
$$
\balpha^\vee_i={2\over\balpha_i^2}\balpha_i.
\efr
With the correct normalization
$$
{\bfmath g}={1\over 2e}\balpha^\vee,\ \ \ \ 
\balpha^\vee\in
\Lambda_R^\vee.
\efr 
The vectors $\lambda\balpha_i^\vee$, where $\lambda$ is a possible
normalizing factor, are the simple roots of a Lie algebra $g^\vee$,
corresponding to a group $G^\vee$ [\Ref{GNO}],
the `dual group' to $G$. All the simply-laced and 
exceptional groups are self-dual. The only non-self dual groups are
${\rm SO}(2n+1)\leftrightarrow{\rm Sp}(n)$.

The magnetic charge of such a solution is defined by
$$
Q_m=-\vert{\bfmath v}\vert^{-1}\int dS_i\,{\rm Tr}\left(B_i\phi\right)=
4\pi{\bfmath g}\cdot\hat{\bfmath v}, 
\efr
where the integration is over the surface of the sphere at infinity.

In [\Ref{WB1}] Weinberg, following Bais [\Ref{BAIS}],
constructs a number of BPS monopole solutions which are spherically
symmetric in the radial gauge.\note{The
construction of monopole solutions for arbitrary gauge groups has also
been considered in [\Ref{MONM}].}
Let $\phi^k(x;\lambda)$ and $A_i^k(x;\lambda)$ be the SU(2)
Prasad-Sommerfield monopole or anti-monopole solution [\Ref{PS}] 
corresponding to
the Higgs expectation value at infinity being $\lambda$.\note{With
respect to an 
anti-hermitian basis $t^k$, $k=1,2,3$, with $[t^j,t^k]=-\epsilon_{jkl}t^l$.}

To construct Weinberg's spherically symmetric solutions
take a homomorphism of SU(2) 
into $G$ defined by a root $\balpha$, with
$\balpha\cdot{\bfmath v}\neq0$ (i.e. $\balpha$ is {\it not\/} a root
of $k$):
$$
\eqalign{
t^1(\balpha)&=(i/2)\left(E_\balpha +E_{-\balpha }
\right),\cr
t^2(\balpha)&=(1/2)\left(E_\balpha -
E_{-\balpha }\right),\cr
t^3(\balpha)&=i{\balpha \cdot{\bfmath H}\over\balpha^2}.\cr}
\nfr{SHOM}
The monopole (or anti-monopole) solution is then 
$$
\eqalign{
\phi(x)&=\sum_{k=1}^3\phi^k(x;{\bfmath v}\cdot\balpha)t^k(\balpha)+i
\left({\bfmath v}-(\balpha \cdot{\bfmath v}/\balpha ^2)
\balpha \right)\cdot{\bfmath H},\cr
A_i(x)&=\sum_{k=1}^3A_i^k(x;{\bfmath v}\cdot\balpha)t^k(\balpha).\cr}
\efr
By expanding these solutions in the asymptotic regime we find they have
${\bfmath g}=1/(2e)\balpha^\vee$ and so
magnetic charge
$Q_m=(2\pi/e)\balpha^\vee\cdot\hat{\bfmath v}$.
Their masses saturate the Bogomol'nyi bound:
$$
\tilde M_\balpha ={2\pi\over e}
\left\vert\balpha^\vee\cdot{\bfmath v}\right\vert.
\efr

By this construction, it appears that fundamental
monopole or anti-monopole solutions are associated to any root of
the algebra $g$ with non-zero inner product with $\bfmath v$.
This is not true for three reasons. Firstly, some of the
solutions may actually consist of superpositions of 
several stationary fundamental 
monopoles at the same point. Secondly, there are more SU(2)
homomorphisms into $G$ than suggested by \SHOM. In fact we shall find that
they sweep out moduli spaces of which \SHOM\ is just a
point. Finally, it is known that for non-simply-laced groups
not all the monopole solutions can be understood in terms of
embeddings of the SU(2) monopole [\Ref{WB2},\Ref{WB4}].
 
Turning to the first point, the question as to whether the
solutions above are fundamental can be answered by calculating the
number of zero-modes around a solution. 
If a solution is not fundamental then there will exist 
additional zero-modes which describe the freedom to
alter the relative separation of the fundamental monopoles.
Using an index theory calculation,
Weinberg [\Ref{WB1},\Ref{WB2},\Ref{WB3}]
argued convincingly that non-fundamental solutions are associated to those
roots $\bgamma$ such that
$\bgamma^\vee=\balpha^\vee+\bbeta^\vee$ where $\balpha$ and $\bbeta$ are
either both positive roots or negative roots of $g$, i.e.
$\tilde M_\bgamma=\tilde M_\balpha+\tilde M_\bbeta$. In other words
these non-fundamental solutions are simply superpositions of
stationary fundamental monopoles.

We now investigate the second point relating to the SU(2)
homomorphisms into $G$. Notice, that ${\bfmath
t}(\balpha)=(t^1(\balpha),t^2(\balpha),t^3(\balpha))$,  
the SU(2) related to the root
$\balpha$ in \SHOM, 
can be deformed by conjugation with constant elements of $H$,
the unbroken gauge group, to give a continuous family
of degenerate monopole solutions:
$$
{\bfmath t}\rightarrow\left[{\rm Ad}\,U\right]{\bfmath t},\ \ \ U\in H,
\nfr{TRA}
where $[{\rm Ad}\,U]x=UxU^{-1}$.
Let us identify this family of solutions. 
A certain subgroup $H_0\subset H$ will
leave the SU(2) invariant, and so the space of SU(2) homomorphisms
connected to ${\bfmath t}(\balpha)$ will be a quotient space 
$H/H_0$ defined by identifying elements of $H$ under right action by
elements of $H_0$. The stabilizer of
${\bfmath t}(\balpha)$, 
$H_0$ has the form $H_0=\U(1)^{r'}\times K_0$.
The $K_0$ part of $H_0$ acts only on the
non-abelian part of $H$ and $r'-1$ of the abelian parts of $H_0$ act
only on the abelian parts of $H$. Precisely one of the $\U(1)$ factors
of $H_0$ acts non-trivially on both the abelian part and non-abelian
parts of $H$. Hence, up to a finite subgroup, 
the quotient $H/H_0$ has the form
$[\U(1)\times(K/K_0)]/\U(1)$ which can be described
locally as the product
$$
{\rm U}(1)\times[K/K_0\times\U(1)].
\nfr{QS}
Globally the quotient is a fibre bundle with base space $K/K_0\times\U(1)$ 
and fibre $\U(1)$. In general, if the finite subgroup 
$Z$ is not contained within $\U(1)\times K_0$ there will
be a residual $Z$ action which must be divided out.

The $\U(1)$ action given by the fibre corresponds to transformations 
of the form 
$$
t^3(\balpha)\rightarrow t^3(\balpha),\ \ E_{\pm\balpha}\rightarrow\exp(\pm 
i\theta)E_{\pm\balpha},
\efr
generated by $i{\bfmath v}\cdot{\bfmath H}$.
These are the charge rotations familiar from the SU(2) monopole 
[\Ref{JZ}]. The additional factor 
${\cal M}=K/K_0\times\U(1)$, 
forming the base space of the bundle, is the space swept out by transformations
generated by step operators $E_\bbeta$ of $k$ with $\bbeta\cdot\balpha\neq0$. 
In general, this part which arises from the non-abelian part of $H$, is
the product of several Lie group coset spaces, a fact which will
play an important role in our subsequent analysis.

The family of solutions swept out from an SU(2) homomorphism defined by
the root $\balpha$ can contain the solutions defined by other
roots which differ from $\balpha$ by roots of the Lie algebra of 
$K$. These solutions are
connected as in \TRA\ by elements of the Weyl group of $K$ 
[\Ref{WB1}]. From our perspective, there is actually nothing
special about the solutions corresponding to a particular root, any
homomorphism of SU(2) into $G$ specified by a point in the space \QS\ 
is equally as good.

These results suggest that the moduli space of a fundamental monopole
will be of the form which is locally 
${\Bbb R}^3\times\U(1)\times{\cal M}$ where the
first factor represents the centre-of-mass position of the monopole
and the second factor corresponds to U(1) charge rotations. The
additional factor $\cal M$, which arises when the unbroken gauge group is
non-abelian, reflects the coset degeneracy of solutions described
above. However, as we discussed in the introduction, there are
certain well-known problems with this picture for the case of a single
monopole in isolation. Before we review these problems and our
proposed solution, we return to the question of 
whether all monopole solutions can be understood in terms of
embeddings of the SU(2) monopole. In fact it is known that
in certain cases with non-simply-laced gauge groups,
when the monopoles are constructed from 
short roots, this is not the case. In these situations,
an `accidental' degeneracy can occur with
two monopole solutions corresponding to two different length roots have
the same mass but whose associated SU(2) homomorphisms 
are not connected by conjugation with an element of the
Weyl group of $K$. Weinberg showed for the case when 
$G={\rm SO}(5)$ [\Ref{WB4}] (see also [\Ref{A1}]) 
that there is a whole family of interpolating solutions
between these degenerate monopoles. 
These additional solutions are manifested by the existence
of zero-modes and collective coordinates
which are apparently not related to any symmetry. 

\chapter{Monopole zero-modes}

In this section we consider the various zero-modes of a single
monopole in a non-degenerate situation (see the remark at the end of
the last section). To each degree-of-freedom of a monopole solution we
can associate a collective coordinate $X^a$.  Variations of the
collective coordinates lead to small fluctuations of the fields
$\delta_aA_i$ and $\delta_a\phi$ which automatically satisfy the 
linearized equations-of-motion.
However, in general they do not satisfy
the back-ground gauge condition:
$$
D_i\delta_aA_i+[{\rm ad}\,\phi]\delta_a\phi=0,
\nfr{GAU}
where $[{\rm ad}\,\phi]\psi=[\phi,\psi]$, which ensures that they are
orthogonal to local gauge transformations. The physical 
zero-modes are equal to
the variations of the solutions by the collective coordinates and a
compensating gauge transformation which ensures that the gauge
condition \GAU\ is satisfied:
$$
\eqalign{
\delta_aA_i&={\delta A_i(X)\over\delta X^a}+D_i\epsilon_a,\cr
\delta_a\phi&={\delta\phi(X)\over\delta X^a}+[{\rm ad}\,\phi]\epsilon_a.\cr}
\efr
Thus $\epsilon_a$ is the generator of a local gauge transformation
which must tend to zero at infinity.

To perform a semi-classical quantization one needs to calculate the
moments-of-inertia tensor corresponding to the zero-modes which
defines a Riemannian metric on the moduli space:
$$
{\cal G}_{ab}
=-\int d^3x\,\left({\rm Tr}\left[\delta_aA_i\delta_bA_i\right]+
{\rm Tr}\left[\delta_a\phi\delta_b\phi\right]\right).
\efr

We are interested in finding zero-modes associated with variations of
the coordinates which parametrize the space of SU(2) homomorphisms. 
Such zero-modes will be generated by gauge transformations with parameters
$\Omega_a$ which must approach a constant at infinity [\Ref{A3},\Ref{A2}]:
$$
\delta_aA_i=D_i\Omega_a,\ \ \delta_a\phi=[\phi,\Omega_a],
\efr
which in unitary gauge outside the monopole core take values
in the Lie algebra $h$ of the unbroken gauge group $H$. 
In order that they be zero-modes we
require that they satisfy the gauge condition \GAU\
which implies the following Laplace-like equation for the gauge parameters:
$$
D_iD_i\Omega_a+[{\rm ad}\,\phi]^2\Omega_a=0.
\nfr{LE}

In order to determine which gauge transformations lead to zero-modes
consider the solution of \LE.
The only non-trivial solutions correspond to
generators of $h$ which fail to commute with ${\bfmath t}(\balpha)$. These are
$t^3(\balpha)$ itself and the step generators $E_\bbeta$ of the Lie
algebra $k$ with
$\bbeta\cdot\balpha\neq0$. For non-degenerate cases $\balpha$ is a
long root and the inner product
can be $\bbeta\cdot\balpha=\pm\balpha^2/2$ only.
The element $t^3(\balpha)$ generates the U(1)
charge rotations whilst the step generators of $k$ can be thought of
as being associated to tangent vectors of the coset space
$\cal M$. Outside the monopole core, we can expand
the gauge parameter in terms of these generators:
$$
\Omega(x_i)=\sum_\bbeta\Omega_\bbeta(x_i)E_\bbeta+\Omega_0(x_i)t^3(\balpha).
\efr
The possible solutions for each of the components $\Omega_\bbeta(x_i)$ 
are written in terms of monopole harmonics [\Ref{WY}]:
$$
\Omega_\bbeta(r,\theta,\phi)=r^\alpha Y_{qlm}(\theta,\phi),
\efr
where $q=-\balpha\cdot\bbeta/\balpha^2=\pm1/2$ is the eigenvalue of
$[{\rm ad}\,it^3(\balpha)]$ 
on the generator $E_\bbeta$. The quantum numbers $l$
amd $m$ are the total angular momentum and its $x_3$ component, so that 
$l=0,1/2,1,\ldots$ and $-l\leq m\leq l$. The allowed values of $l$ are $\vert
q\vert,\ \vert q\vert+1,\ldots$. 
The parameter $\alpha$ is determined from the radial equation:
$\alpha(\alpha+1)=l(l+1)$. 

The gauge parameter corresponding to the abelian subgroup of $H$
generated by $t^3(\balpha)$ is simply a constant
in the region outside the monopole core. This transformation leads to
a normalizable zero-mode of the monopole which corresponds to
the freedom to perform U(1) charge rotations.  
In fact taking into account the
fields in the core, it actually dies off exponentially outside the
core and so its associated moment-of-inertia $\sim R_{\rm core}$.

The non-abelian part of the unbroken gauge group
leads to gauge transformations $\Omega_\bbeta(x_i)$ for each $\bbeta$ with
$\bbeta\cdot\balpha\neq0$. 
In Appendix A, we analyse the corresponding Laplace
equation in the region outside the core. The results specialize those of
Abouelsaood [\Ref{A3},\Ref{A2}] to the case of BPS monopoles. 
We find four solutions with
quantum numbers $(q,l,m)=(\pm\half,\half,\pm\half)$. For each of these
solutions the radial equation dictates that $\alpha=1/2$. The 
gauge parameters $\Omega_{\bbeta}$ therefore grow like ${\cal
O}(r^{1/2})$ which clearly
contradicts our condition that the resulting modes should correspond
to a global gauge rotation of the monopole at large distance. 
Thus the monopole has bosonic zero-modes which die off like
${\cal O}(r^{-1/2})$ at large distance, rather than the expected
${\cal O}(1/r)$, and have no obvious relation to the action of the
global non-abelian symmetry generators. 
For the considerations of the next section it is important
that all of these modes are well-behaved in the core of the monopole. 
To determine this we should solve the linearized Bogomol'nyi equation
and gauge condition at short distance also. Fortunately this has
already been accomplished by Weinberg in Appendix C of [\Ref{WB1}] who
solved an equivalent Dirac equation. Weinberg's analysis reveals the
same number of bosonic zero-modes which go like ${\cal O}(r^{-1/2})$ at
large distance. Each of these modes is well-behaved at $r=0$ as 
required.\note{Explicitly, the bosonic modes
which we have found are related by his equation (3.5) to the Dirac
zero-modes implied by his equation (C.14) where we set $a_{1}=0$.   
Note that Weinberg works in radial gauge 
where the monopole has spherical symmetry.}

\chapter{Multi-monopole solutions and the non-abelian zero-modes}

In the last section we have seen that there is a problem in proceeding
to a semi-classical quantization of a monopole when it transforms
under the action of the non-abelian unbroken symmetry group. The usual
relation between continuous degeneracies of classical solutions 
and normalizable zero-modes of the small fluctuation operator breaks
down. This relation is the cornerstone of the usual method of
collective coordinates and without it, it is not clear how to proceed.   
Before we can make progress, it is clear 
that we need some way of regulating the
pathological behaviour of the zero-modes identified above. Fortunately
such a scheme is available. Coleman and Nelson [\Ref{NC}] consider a 
chromo-magnetic dipole, which consists of a separated monopole ($M$) and an
anti-monopole ($\bar M$). They work in an SU(3) theory with a non-zero scalar
potential and therefore outside the BPS limit. 
The fields of $M$ and $\bar M$ are chosen 
so that at long distance the gauge potential falls off faster than the
${\cal O}(1/r)$ of a single
monopole. This requires that $M$ and $\bar M$ are
characterized by elements ${\cal G}_\pm$ with ${\cal G}_++{\cal
G}_-=0$, see eqn \ASY. Away from the BPS limit, the Higgs fields fall
off to their vacuum expectation value 
exponentially and can be ignored at long distance.
In the dipole system, there {\it are\/} now zero-modes which correspond to
global colour rotations which have a finite norm growing like $R$, the
separation of the sources. These modes correspond to gauge transformations
which approach a constant at distances much greater than $R$.
The important point is that
they are not localized around the cores, since the gauge
parameters fall off as $(r/R)^{\sqrt{3/4}-1/2}$ in the vicinity of 
$M$ or $\bar M$, which vanishes for any fixed $r$ as
$R\rightarrow\infty$. This result is slightly different 
from the $r^{1/2}$ behaviour found in the last section;  
this is because in the BPS
limit the Higgs field contributes to the behaviour of the modes as
indicated in \LE. The non-abelian global colour modes are therefore 
interpreted as being a property of the non-abelian flux tube
joining the two sources. A semi-classical quantization can now be
performed. The non-abelian degrees-of-freedom lead to quantum mechanics on the
coset space $\SU(2)/\U(1)$.\note{This space is even simpler than the
one considered in [\Ref{NC}] because we are suppressing additional
$\U(1)$ factors which correspond to independent electric charge
rotations of $M$ and $\bar{M}$.} The states of this quantum mechanics are
are chromo-dyons 
associated to representations of $\SU(2)$ and have energies $\sim
e^2/R$. Coleman and Nelson go on to argue that these diffuse
excitations of the colour flux tube effectively 
disappear as $R\rightarrow\infty$ and
one is left with only the ground-state of the quantum mechanics. 

We would like to understand the issue of global colour transformations
for a theory with a more general gauge group and specifically 
in the BPS limit. Unfortunately, we immediately run into a
problem because although we can arrange for the long-range parts of
the gauge fields of $M$ and $\bar M$ to cancel, the ${\cal O}(1/r)$ 
parts of their Higgs fields would
add. When we introduce extended supersymmetry the situation is even 
more inconvenient because the $M-\bar M$
configuration is not invariant under any of the
supersymmetry generators since it does not saturate the Bogomol'nyi bound. 
A moment's thought shows that the problem of the
non-abelian modes is intimately related with the fact that the
corresponding gauge transformations do not commute with the long-range
${\cal O}(1/r)$ part of the non-abelian gauge field and Higgs 
field; the part lying inside the Lie
algebra $k$. So rather than than consider a configuration
where all the ${\cal O}(1/r)$ parts of the gauge field and Higgs field 
are cancelled, we can
construct a configuration consisting purely
of monopoles (i.e. no anti-monopoles), where
only the non-abelian ${\cal O}(1/r)$ parts of the fields 
are cancelled. 

To implement this idea, consider a configuration of monopoles defined
by elements ${\cal G}_i$. In order to superimpose these solutions to
get a multi-monopole solution we have to ensure that $[{\cal
G}_i,{\cal G}_j]=0$. We then choose the first monopole to have
arbitrary magnetic charge. For each possible choice of the first
monopole it is always possible to find a set monopoles so that the
${\cal O}(1/r)$ non-abelian parts of the long range field are
cancelled. This requires that
$\sum_i{\cal G}_i$ lies entirely within the abelian part of $h$.
In cases involving groups
with real representations a single extra monopole will suffice to
cancel the long-range non-abelian field, however, in the cases with
complex representations it will be necessary to have more than one
additional monopole. It will be convenient for the following analysis
to have the additional monopole(s) ${\cal G}_i$, for $i>1$, all located
at the same point a distance $R$ from the first monopole.
The conglomerate consisting of the monopoles with $i>1$ will have
its own set of zero-modes, for instance in the case 
when it does consist of more
than one fundamental monopole there will be zero-modes
corresponding to the freedom to separate the monopoles. However, we
are interested in the additional zero-modes that arise in the presence
of the first monopole. 

As we have explained in the last section there are no zero-modes
corresponding to non-abelian global colour transformations for a
single monopole. However, 
if we have cancelled the ${\cal O}(1/r)$ non-abelian parts of the field by
adding a set of compensating monopoles some distance away,
then there are normalizable zero-modes corresponding
to global colour rotations {\it of the system as a whole}. 
In such a background 
field configuration there exist solutions of equation \LE\ which
asymptote to a constant as required.
The argument is directly analogous to the argument of
Coleman and Nelson for the monopole anti-monopole system 
(see the Appendix of [\Ref{NC}]). The problem of solving the linear 
partial differential equation \LE\ with constant boundary conditions
is equivalent to the problem of inverting the differential operator
which occurs on the LHS of the equation in an appropriate space of
functions. These authors show that this problem is in turn
equivalent to the problem of minimizing a differentiable, convex,
function which is bounded below. The existence of a solution to this
latter problem is then guaranteed by standard theorems from the 
calculus of variations. The whole argument depends only on the large
distance fall-off of the fields which contribute to the LHS of \LE\ . 
In the present case, there is a power-law contribution to the Laplace
equation from the Higgs field. However, its fall-off far away from the
sources is of the same form as the gauge field ${\cal O}(1/r^{2})$. 
In addition, the fact that there are still 
long-range abelian fields in our system does not affect the result,
because the non-abelian generators of $k$ commute with these abelian fields.
Each pair of generators $E_{\pm\bbeta}$ of $k$ such that 
$\bbeta\cdot\balpha\neq0$ will lead to two non-abelian modes of the
combined system. In the vicinity of 
the monopole, the two normalizable modes must be a
certain linear combinations of the four modes that we discussed in the
previous section. In Appendix A we determine the unique linear
combinations with the correct angular behaviour. 
On dimensional grounds the modes are generated by gauge
parameters which behave as
$(r/R)^{1/2}$ near the monopole. 
As pointed out by Coleman and Nelson as
$R\rightarrow\infty$ they are `expelled' from the core. 

To summarize, in the combined system, there will be an additional
normalizable non-abelian zero-modes which correspond to global colour
transformations. There will be one such mode for each direction in the
coset $\cal M$. By a straightforward extension of the arguments given in  
[\Ref{NC}] these modes have a norm which grows linearly with $R$, the 
separation of the monopole and the conglomerate. 

If we now perform a semi-classical quantization for the
degrees-of-freedom associated to the non-abelian zero-modes of the
combined system then we will be led to consider quantum mechanics on
the coset space $\cal M$. States will be associated to
representations of $K$ and have energies determined by the scale
$e^2/R$. Our interpretation 
of these chromo-dyon states is exactly the same as that of 
Coleman and Nelson. They
correspond to excitations of the non-abelian flux tube joining the
two sources. The ground-state of the quantum mechanics is simply the constant
wavefunction which is a singlet under the non-abelian group $K$.  

\chapter{Supersymmetry and evidence for GNO duality}

In this section we consider the generalization of our analysis of the 
multi-monopole system to theories with extended supersymmetry.   
The bosonic zero-modes of a multi-monopole configuration 
come in multiples of four.
The reason is that the Bogomol'nyi equations can be
formulated as the self-dual Yang-Mills equations for a
time-independent
Euclidean gauge field with components $W_m=(A_i,\phi)$, with
$m=1,2,3,4$. This auxiliary
Euclidean space has three 
inequivalent almost complex structure $J_{mn}^{(i)}$, $i=1,2,3$,
since it is a hyper-K\"ahler manifold. If we have one bosonic mode
$\delta W_m$ which is a solution of the linearized self-dual
Yang-Mills equations
then by acting with the three almost complex structures we can
generate three other modes $J_{mn}^{(i)}\delta W_n$
(see for example [\Ref{G1}]). What is not clear from this analysis 
is the exact relation between these zero-modes and 
corresponding infinitesimal changes of the collective coordinates 
of the background field configuration. For the charge rotation zero-mode of
a single monopole, there is no difficulty in interpreting the other
three zero-modes. They correspond to moving the centre-of-mass of the
monopole. For the non-abelian global colour 
modes of the multi-monopole solution
the situation is more subtle. In this case the non-abelian 
modes come in multiples
of two for each positive root of the Lie algebra $k$ with 
$\bbeta\cdot\balpha\neq0$. We have argued in Appendix A that
these two modes are related by just one of the almost complex
structures which we denote as $J_{mn}$. A specific almost complex structure
in spacetime is picked out simply because our system has a preferred
direction, the axis between the monopole and the conglomerate. Its
explicit form is given in the Appendix but plays no role in the
following discussion. The interpretation of the
other two zero-modes generated by the action of the two other almost
complex structures is not so clear. The question is whether they
correspond to variations of some collective coordinates of the
configuration? As we have mentioned in the introduction, the only
reasonable candidates for these coordinates are the relative non-abelian
group orientation of the monopole and the conglomerate. These
correspond to functional directions which take us outside the space of
configurations for which the long-range non-abelian fields cancel.
They certainly do not in any case correspond to global gauge
transformations. We we will assume in the rest of our analysis that
these modes do not affect our arguments regarding the global colour
zero-modes. 

Let us choose some basis for the global colour zero-modes $\delta_aW_m$, where
$a=1,\ldots,{\rm dim}({\cal M})$. The single almost complex structure
$J_{mn}$ induces an action on these zero-modes
$$
{\cal J}_{ab}\delta_bW_m=-J_{mn}\delta_aW_n,
\efr
which naturally interchanges the pair of modes associated to the
generators $E_{\pm\bbeta}$.
The moments-of-inertia tensor of the non-abelian modes defines a 
metric on $\cal M$
$$
{\cal G}_{ab}=-{1\over R}\int d^3x\,{\rm
Tr}\left(\delta_aW_n\delta_bW_n\right),
\nfr{MET}
where we have separated out the overall scale $R$ to make the metric
dimensionless. This metric is
the natural $K$-invariant metric on the coset space $\cal
M$. Furthermore, the almost complex structure defined previously
descends to the almost complex structure ${\cal J}_{ab}$ 
on $\cal M$. (Recall that
$\cal M$ is a K\"ahler manifold and so admits one almost complex structure.)

Consider to begin with an $N=2$ supersymmetric gauge theory.\note{Our
notations follow exactly those of Gauntlett [\Ref{G1}] except that we
use $\mu$ etc. for spacetime indices and $m$ etc. as indices of the
auxiliary Euclidean space.} Our discussion will be brief because the
necessary details of the semi-classical quantization in such theories 
has appeared elsewhere. Our approach follows very closely the
approach adopted in [\Ref{G1},\Ref{BL},\Ref{HS}].
For an $N=2$ theory there is single Dirac spinor which
is the super-partner of the Euclidean gauge field. We now have to consider 
possible fermionic zero-modes that can arise. 
In the background of a set of monopoles
which saturate the Bogomol'nyi bound
half the supersymmetries of the theory are broken. These are the
anti-chiral supersymmetry transformations in the Euclidean
space. The zero-modes of
the configuration naturally form multiplets of the unbroken
supersymmetry generated by chiral spinors. Usually this pairs
the four bosonic zero-modes related by the three almost complex structures
with two fermionic zero-modes. For the non-abelian zero-modes this is
no longer the case. Instead two non-abelian bosonic zero-modes are
naturally paired with one fermion zero-mode by
half of the unbroken supersymmetries. 
To make this explicit, we define following [\Ref{G1},\Ref{HS}] 
a $c$-number chiral (Dirac) spinor satisfying
$$
\epsilon_+^\dagger\epsilon_+=1,\ \ \
J_{mn}\Gamma_n\epsilon_+=i\Gamma_m\epsilon_+.
\efr
where the $\Gamma_{n}$ are hermitian, Euclidean gamma matrices.  
The fermionic modes pair with the normalizable bosonic modes
$\delta_aW_m$ as
$$
\chi_a=\delta_aW_m\Gamma_m\epsilon_+.
\efr
However, by construction, only half of these fermion zero-modes 
are independent since 
$$
{\cal J}_{ab}\chi_b=i\chi_a.
\nfr{FP}
One important property of the fermion zero-modes associated
to the non-abelian bosonic zero-modes is that 
they carry zero space-time spin. This follows from the fact that,
unlike the more familiar fermion zero-modes which are paired with the
spacetime and charge degrees-of-freedom of the monopole, these modes 
transform in the fundamental representation of the $\SU(2)$ subgroup in
which the monopole is embedded. 

The next stage in the semi-classical quantization program is to
perform an expansion of the action in terms of collective coordinates
[\Ref{G1},\Ref{HS}]. First of all, we introduce a set of time-dependent
coordinates $X^a$ on
the coset space $\cal M$. For the $N=2$ theory, each bosonic coordinate
$X^a$ of the coset space is
accompanied by a fermionic coordinate $\lambda^a$. Because of \FP\ the
fermionic coordinates are not independent. They are
related by the complex structure ${\cal J}_{ab}$ of $\cal M$:
$$
-i\lambda^a{\cal J}_{ab}=\lambda^b.
\efr
The resulting quantum mechanics that arises from the non-abelian
degrees-of-freedom is described by the action
$$
S_{\rm eff}={R\over e^2}\int dt\,{\cal G}_{ab}\left[\dot X^a\dot X^b
+4i\lambda^{\dagger a}D_t\lambda^b\right].
\efr
In the above, ${\cal G}_{ab}$ is the $K$ invariant 
metric on the coset space $\cal M$ defined in \MET.
The quantum mechanics is $N=4\times\half$ supersymmetric. The
supersymmetry transformations are explicitly
$$\eqalign{
\delta X^a&=i\beta_1\lambda^a+i\beta_2\lambda^b{\cal J}_{ba}\cr
\delta\lambda^a&=-\beta_1\dot X^a-\beta_2\dot X^b{\cal J}_{ba},\cr}
\efr
where $\beta_1$ and $\beta_2$ are two real parameters. It is
well-known [\Ref{AG}] that states in this theory are in one-to-one
correspondence with the holomorphic forms on ${\cal M}$. Excited
states of the theory will have energies $\sim e^2/R$. 
Our interpretation of these states is exactly
analogous to the situation in the purely bosonic theory. States with
non-zero energy are identified with excitations of the non-abelian
colour flux tube linking the two sources. Zero energy states, on the
other hand will be interpreted as being due to a degeneracy of the
monopole itself. Ground-states of the quantum mechanics correspond to
the Dolbeault complex of holomorphic 
harmonic forms on $\cal M$, i.e. associated to elements of the
cohomology $H^{p,0}({\cal M})$. For the coset spaces we are consider only
$b^{p,p}\neq0$ and in addition $b^{0,0}=1$, and so there is only a single 
ground-state corresponding to the constant function. Hence in the
context of an $N=2$ supersymmetric gauge theory (with no matter
fields) the conclusion is the
same as for a purely bosonic gauge theory: the monopoles carry no
additional degrees-of-freedom.

The situation is different in an $N=4$ supersymmetric gauge
theory. Now there are two Dirac fermion fields which means that the
number of fermion zero-modes is doubled. To each bosonic coordinate of
the space $\cal M$ there is an associated two component real spinor
$\psi^a$. The resulting supersymmetric quantum mechanics is described
by an action [\Ref{BL}]
$$
S_{\rm eff}={R\over e^2}\int dt\left[{\cal G}_{ab}\dot X^a\dot
X^b+i{\cal G}_{ab}\bar\psi^a\gamma^0D_t\psi^b+{1\over6}{\cal R}_{abcd}
\bar\psi^a\psi^c\bar\psi^b\psi^d\right].
\efr
In the above ${\cal R}_{abcd}$ is the Riemann tensor of $\cal M$. The
quantum mechanics now has an $N=2\times 1$ supersymmetry. It is
well-known [\Ref{AG}] 
that the states of the theory are in one-to-one
correspondence with the forms on $\cal M$. As before, we are interested
in the number of ground-states of the quantum mechanics which in this
case correspond to the de Rahm complex of
harmonic forms on $\cal M$. For the coset spaces we are considering the
non-trivial cohomology appears in the $H^{p,p}({\cal M})$. So all the
ground-states are bosonic since they correspond to forms of even
degree and the degeneracy is equal to the Euler character $\chi({\cal
M})$. Furthermore, the harmonic forms
are precisely the $K$-invariant forms [\Ref{HEL}].

As described previously, the
states with non-zero energy are interpreted as being excitations of the
colour flux tube connecting the two sources.
Our interpretation of the degenerate ground-states is rather
different. Their energy, being zero, 
is manifestly independent of the separation of the sources so
they represent a true 
degeneracy of the combined system. Since the
combined system is constrained to have vanishing ${\cal O}(1/r)$ non-abelian
gauge and Higgs fields this degeneracy is the same as the degeneracy of the
single monopole associated to $\balpha$. With GNO duality in mind,
the degenerate ground-states have three important properties: 
they are all bosonic; they carry zero space-time spin; they
are all singlets of the unbroken gauge group. 

In the subsequent section we shall find by example (at least for
non-degenerate cases) that this degeneracy is exactly what is required
for the monopoles to fill out the representations of the dual gauge
group. So we have found very strong evidence of GNO
duality the context of $N=4$ supersymmetry. 
Interestingly the dual gauge group acts on the
space harmonic forms on the manifold $\cal M$ which suggest that the
generators of the dual gauge group should somehow be realized as bilinears of
the fermion zero-modes.

\chapter{Examples}

In this section we discuss a number of examples which allows us to
illustrate that the multiplicity of monopoles is consistent with the
GNO duality conjecture. We also discuss the cases
with degenerate monopole solutions (the case when monopole solutions have the
same mass and magnetic charge, but which are not related by
conjugation in $H$) where the analysis is more subtle due to the
existence of additional zero-modes.

The space $\cal M$ is in general the product of several Lie algebra
coset spaces. For convenience we list all the spaces that appear in our
construction and their Euler characters below:
$$\eqalign{
\SU(n+m)/\SU(n)\times\SU(m)\times\U(1),&\ \ \ (n+m)!/(n!m!),\cr
{\rm SO}(2n)/\U(n),&\ \ \ 2^{n-1},\cr
{\rm SO}(n+2)/{\rm SO}(n)\times\U(1),&\ \ \ n+1,\ n\in2{\Bbb Z}+1;
\ n+2,\ n\in2{\Bbb Z},\cr
{\rm Sp}(n)/\U(n),&\ \ \ 2n,\cr
E_6/{\rm SO}(10)\times\U(1),&\ \ \ 27,\cr
E_7/E_6\times\U(1),&\ \ \ 56,\cr
{\rm Sp}(n)/{\rm Sp}(n-1)\times\U(1),&\ \ 2n.\cr}
\efr
All these spaces except the last are hermitian symmetric spaces.
For the simply-laced cases the Euler character is simply equal to the
dimension of the representation associated in the normal way to spot
of the Dynkin diagram of the group in the numerator of the coset that must be
removed to give the Dynkin diagram of the group in the denominator of
the coset.\note{For a discussion of the cohomology of these spaces in
the context of conformal field theory see
[\Ref{LVW1}]. For a discussion in the mathematical literature see
[\Ref{BOTT}].} 

(i) $G={\rm SU}(7)$. Choose the Higgs field so that $v_1$, $v_2$,
and $v_5$ are non zero. The unbroken gauge group is
$[{\rm U}(1)^3\times\SU(3)\times\SU(2)]/Z$. The stable massive gauge bosons
form multiplets of $K$ with lowest weights given by 
the simple roots $\balpha_i$, $i=1,2,5$. So under $H$ the
quantum numbes of the representations are
$$
\balpha_1:\ \ (0,0)_{v_1},\ \ \ 
\balpha_2:\ \ (\bar 3,0)_{v_2},\ \ \ 
\balpha_5:\ \ (3,2)_{v_5},
\nfr{MUL}
where the subscript is the electric charge  of the multiplet in units of $e$. 
The negatives of these roots give the complex conjugate representations.

Fundamental monopole solutions can be associated to the roots
$\balpha_i$, $i=1,2,5$. 
The non-abelian modes are associated to the coset spaces
$$\eqalign{
&\balpha_1:\ \ 1,\cr
&\balpha_2:\ \ \SU(3)/\SU(2)\times\U(1),\cr
&\balpha_5:\ \ [\SU(3)/\SU(2)\times\U(1)]\times
[\SU(2)/\U(1)],\cr}
\efr
Notice that the factors in these moduli spaces are hermitian symmetric
spaces: $\SU(3)/\SU(2)\times\U(1)\simeq{\Bbb C}P^2$ and
$\SU(2)/\U(1)\simeq{\Bbb C}P^1$. Following the discussion in section
7 the degeneracy of these monopole solutions is given by the Euler
characteristic of the coset manifolds and so we find multiplicities and
magnetic charges (in units of $2\pi/e$)
$$
\balpha_1:\ \ 1_{v_1},\ \ \ 
\balpha_2:\ \ 3_{v_2},\ \ \ 
\balpha_5:\ \ (3\times2)_{v_5},
\efr
These multiplicities match exactly those in \MUL\ and are consistent
with the GNO conjecture since in this case the group is self-dual.

The generalization to arbitrary SU$(n)$ with arbitrary symmetry
breaking is now obvious. If the unbroken gauge group is $H={\rm
S}\left(\U(n_1)\times\cdots\times\U(n_p)\right)/Z$ then the stable massive
gauge bosons come in multiplets with lowest weights $\balpha_i$ with $i=1+
\sum_{j=1}^{k-1}n_j$ and $n_k=1$, so that $\balpha_i\cdot{\bfmath
v}\neq0$. The gauge bosons associated to this root transforms as a
$(n_{k-1},\bar n_{k+1})$ of $\SU(n_{k-1})\times\SU(n_{k+1})$, with
electric charge $\balpha_i\cdot{\bfmath v}$ and is a
singlet of all the other factors in $H$. The negative simple root
$-\balpha_i$ gives a multiplet transforming in the complex conjugate
representation. Monopole solutions are
associated to $\balpha_i$ and the associated non-abelian
degrees-of-freedom are described by the coset space
$$
[\SU(n_{k-1})/\SU(n_{k-1}-1)\times\U(1)]\times
[\SU(n_{k+1})/\SU(n_{k+1}-1)\times\U(1)].
\efr
A factor of the form
$\SU(n)/\SU(n-1)\times\U(1)\simeq{\Bbb C}P^{n-1}$ has Euler character
$n$. So the multiplicity and magnetic charge of the monopole and
anti-monopole are
$(n_{k-1}\times n_{k+1})_{\pm\balpha_i\cdot{\bfmath v}}$.
It is apparent
that the multiplicity of
monopole solutions is consistent with the fact that the gauge group is
self-dual in this case.

(ii) $G=E_6$. Our labelling of the roots of $E_6$ matches that in
[\Ref{CORN}]. We
choose the Higgs vacuum expectation value 
to have $v_2\neq0$. The unbroken gauge group is
$H=[\U(1)\times\SU(2)\times\SU(5)]/Z$. The stable massive gauge bosons
come in multiplets with lowest weights $\pm\balpha_2$ giving
representations $(2,10)_{\pm v_2}$ of $H$.

The monopole solutions are associated to the roots $\pm\balpha_2$
and the non-abelian degrees-of-freedom are described by the coset space
$$
[\SU(2)/\U(1)]\times[\SU(5)/\SU(3)\times\SU(2)\times\U(1)].
\efr
In this case the second factor is an example of a complex Grassmanian
with Euler character $10$. Hence the monopole and anti-monopole have a
multiplicity $(2\times10)_{\pm v_1}$ matching the gauge bosons.

(iii) $G={\rm Sp}(r)$.\note{In our conventions the Lie
algebra of ${\rm Sp}(r)$ has rank $r$.} 
Consider the case when only $v_1\neq0$. In
this case the unbroken gauge group is $H=[\U(1)\times{\rm
Sp}(r-1)]/Z$. The stable massive gauge bosons come in multiplets with
lowest weight
$\balpha_1$, i.e. a $(2r-2)_{v_1}$ of
$H$. The negative simple root $-\balpha_1$ corresponds to the complex
conjugate representation $(2r-2)_{-v_1}$. 

In this case the gauge group is not self-dual, rather ${\rm
Sp}(r)^\vee={\rm SO}(2r+1)$. So we need to enumerate the
multiplicities of monopole solutions in a ${\rm SO}(2r+1)$ theory
with a Higgs field which leads to an unbroken gauge group  
$H^\vee=[\U(1)\times{\rm SO}(2r-1)]/Z^\vee$. 
The monopole solutions are associated to the simple root
$\pm\balpha_1$ and the non-abelian degrees-of-freedom are described by
the coset space
$$
{\rm SO}(2r-1)/{\rm SO}(2r-3)\times\U(1).
\efr
This manifold has Euler 
character $2r-2$. So the multiplicities and magnetic charges
of the monopoles are $(2r-2)_{\pm v_1}$. Again we see that the
multiplicities are exactly what is required by the GNO conjecture.

Unfortunately it is not possible to verify the conjecture in the reverse
direction, i.e. comparing the gauge bosons of the ${\rm SO}(2r+1)$ with
the monopoles of the ${\rm Sp}(r)$ theory, since this is a case with
degenerate monopole solutions (see the remark at the end of this section).

For all the cases without degenerate monopole solutions the following
general picture emerges. Let the gauge group be $G$ with simple roots
$\balpha_i$ and the dual gauge group be $G^\vee$ with simple roots
$\lambda\balpha^\vee_i$. The Higgs vacuum expectation values 
in the two dual pictures are specified by the vectors
${\bfmath v}$ and ${\bfmath v}^\vee$,
respectively. In order that the spectrum of the gauge bosons can be
equal to the spectrum of monopoles in the dual picture, and
vice-versa, requires that the coupling constant of the dual theory is
$$
e^\vee={2\pi\over e\lambda}
\efr
and ${\bfmath v}^\vee={\bfmath v}$.
The stable massive gauge bosons in the $G$ theory are associated to
the simple roots $\pm\balpha_i$ with
$\balpha_i\cdot{\bfmath v}\neq0$. If the unbroken gauge group is
$H=[\U(1)^{r'}\times K_1\times\cdots\times K_p]/Z$, where each of the
factors is simple, then the states 
transform in representations $(R_1,\ldots,R_p)_{v_i}$, where $R_a$
is the representation of $K_a$ with lowest weight given by $\balpha_i$
projected into the root space of $K_a$. The root $-\balpha_i$ gives
the complex conjugate representations.

In the dual description, the gauge group breaks to
$H^\vee=[\U(1)^{r'}\times K_1^\vee\times\cdots\times K_p^\vee]/Z^\vee$.
The monopoles of the dual description are associated to the simple roots
$\lambda\balpha_i^\vee$ and the non-abelian degrees-of-freedom are 
described by the product of coset spaces
$$
[K^\vee_1/L_1]\times\cdots\times[K^\vee_p/L_p].
\efr
Each factor $L_a$ is either $K^\vee_a$, itself, in which case the
factor is trivial, or $L_a$ has the form $L_a=\U(1)\times L_a'$, in
which case the factor is a hermitian symmetric space. The multiplicity
of the monopole solution is then the product of the Euler characters
of these spaces. The fact that the multiplicities of the 
gauge bosons of the $G$ theory associated to
$\balpha_i$ and the monopoles of the $G^\vee$ theory associated to 
$\lambda\balpha_i^\vee$ match follows from the fact that in
non-degenerate cases, and so when $\lambda\balpha_i^\vee$ is a long
root,
$$
{\rm dim}\left(R_a\right)=\chi\left(K^\vee_a/L_a\right).
\nfr{REL}

Even in cases with degenerate monopole solutions the na\"\i ve counting
of states is also consistent with the GNO conjecture. The point is that
as it stands the relation \REL\ no longer holds; however there now
exist monopoles of the dual theory which are degenerate with the
$\lambda\balpha_i^\vee$ monopole but which are related to roots of
$G^\vee$ which are not simple. For example, consider the 
${\rm SO}(2r+1)\leftrightarrow{\rm Sp}(r)$ example 
discussed above. The monopole solutions
of the ${\rm Sp}(r)$ theory associated to the short root $\balpha_1$ have a
multiplicity given by the Euler character of the coset space
$$
{\rm Sp}(r-1)/{\rm Sp}(r-2)\times\U(1),
\efr
which is $2r-2$, and this is not enough states to fill out the
vector representation of the dual unbroken gauge group
$[\U(1)\times{\rm SO}(2r-1)]/Z$. However, there is a degenerate monopole
solution corresponding to the long root
$2\balpha_1+2\balpha_2+\cdots+2\balpha_{r-1}+\balpha_r$ of ${\rm Sp}(r)$. This
solution is invariant under conjugation by the unbroken gauge group
has so (na\"\i vely) contributes one additional state that makes the
multiplicity of monopole states $2r-1$, exactly the dimension of the
vector representation of ${\rm SO}(2r-1)$. 
In spite of this, it is not clear whether this
simple analysis is valid because, as we have previously noted, the
monopole solutions in this case have additional normalizable
zero-modes with associated collective coordinates
which can connect the two degenerate monopole solutions [\Ref{WB4}].

\chapter{Discussion}

In this paper we have obtained the semi-classical spectrum of
non-abelian magnetic monopoles in $N=4$ supersymmetric gauge theory
with a general gauge group. Our results indicate that, at least for
the purposes of counting BPS-saturated states, non-abelian monopoles
can be thought of as having a compact internal moduli space ${\cal M}$
which is generated by the action of the unbroken non-abelian symmetry
group. Although this picture is hard to substantiate for a single
monopole in isolation, we have shown that the usual problems
associated with long-range non-abelian fields can be by-passed by
considering suitable multi-monopole configurations. Standard 
semi-classical arguments then lead one to consider a maximally
supersymmetric quantum mechanics on ${\cal M}$. In this framework, 
the wavefunctions of BPS-saturated states correspond 
to harmonic forms on ${\cal M}$ and the multiplicity of these states
is given by the Euler character, $\chi({\cal M})$. In all cases this
simple result yields exactly the degeneracy of states predicted by 
GNO duality, although the analysis may not be reliable in certain
special cases where it is known that additional continuous
degeneracies of BPS monopole solutions exist [\Ref{WB4}]. 

One of the remaining puzzles is the occurrence of additional
normalizable zero modes of the multi-monopole configurations
discussed in this paper. In fact, the presence of these zero-modes 
is in agreement with the usual expectation that the moduli space 
should be a hyper-K\"{a}hler manifold. This is in contrast to the
coset spaces considered above which are, 
in general, merely K\"{a}hler. We have explicitly assumed that the
presence of additional normalizable zero modes does not alter our
conclusions. In this respect we must place a caveat on the results
described above. We cannot rule out the possibility that there are new 
collective coordinates of the multi-monopole configurations which
correspond directly to the additional zero modes.\note {In fact, as
mentioned above, this possibility is realized in the
degenerate cases described in [\Ref{WB4}].} 
The fact that the modes are normalizable 
suggests that the additional directions in the moduli space would correspond to
new configurations with vanishing long-range non-abelian fields. 
This would imply some hitherto undiscovered
continuous degeneracy of the BPS multi-monopole system, {\it even for
monopoles at arbitrarily large separation}. 
If this were the case then each of the 
coset spaces we have identified would be a subspace of a hyper-K\"{a}hler
moduli space of larger dimension and a complete analysis would lead one to
consider the de Rham cohomology of this larger space. In this case we
must be content to assume that the harmonic forms on the coset sub-space
are in one to one correspondence with the harmonic forms on the full
moduli space. However, there are two other possibilities which we
consider more likely. The first is that the additional collective
coordinates are simply not present.
This would mean that the extra zero modes occur only 
at the level of a linearized analysis and do not correspond to any 
finite variations of the fields. The second possibility is that the
additional collective coordinates are present but 
correspond to directions in function
space which take us outside the subspace of configurations for which
the long-range non-abelian fields cancel. In either of these two cases
our semiclassical results should be reliable without further
assumptions. 

Several features of our results are worthy of comment. Firstly we
expect the degeneracy of states discovered here to persist in other
theories where Montonen-Olive duality is expected to be exact for
gauge group $\SU(2)$ [\Ref{GH}]. An
example is $N=2$ super Yang-Mills theory coupled to $N_{f}$ matter
hypermultiplets in the fundamental representation, where $N_{f}$ is
chosen so that the perturbative $\beta$-function vanishes. For $N_f$
less than this critical value, the $N=2$ theories are not in a
non-abelian coulomb phase and there is no non-abelian gauge symmetry
in the infra-red. Theories of this type with general gauge groups
have been considered (see for example [\Ref{NTWO}]).
It seems possible that the form of exact GNO duality 
which appears to hold in the 
$N=4$ case is, however, relevant to the understanding of the effective 
duality in the non-abelian coulomb phase
proposed by Seiberg in $N=1$ theories [\Ref{SEI}].
 
Finally, our analysis provides an explicit semi-classical description of the 
states which make up the multiplets of the dual gauge group. They 
consist of monopoles with different (even) numbers of fermion zero
modes excited. This suggests that it may be possible to find an
explicit realization for the generators of the dual gauge
group in terms of the fermionic fields in the original theory.  

The results given in this paper are incomplete in a number of
respects. In our analysis, we have not considered the spectrum of 
dyons that would result from exciting the abelian charge
degrees-of-freedom of the monopoles. In addition we have set the theta
angle to be zero. It would be interesting to see how the extended
${\rm SL}(2,{\Bbb Z})$ duality, which arises from considering 
non-zero theta angle in the SU(2) theory, appears in the case of a 
general group. These questions are under active investigation. 

We would like to thank Ed Corrigan for useful discussions. ND and TJH
are supported by PPARC Advanced Fellowships. 

{\bf Addendum:} Since completing this work we have recieved two preprints
[\Ref{G2},\Ref{W6}] which
analyse the spectrum of monopoles and dyons in 
N=4 supersymmetric gauge theory in the case of maximal symmetry
breaking. Specifically these authors consider the case of gauge group 
$G=\SU(3)$ breaking down to $H=\U(1)\times \U(1)$. As the  
results pertain to a theory where only an abelian subgroup is left
unbroken they do not overlap directly with ours. However it is
interesting to consider the spectrum of monopole states in their model
in the limit in which the non-abelian symmetry $\SU(2)\times\U(1)$ 
is recovered and compare with our results. Specifically we can compare
to the analysis of the case $\SU(3)\rightarrow \SU(2)\times\U(1)$
given in Section 1 above. 
Using the notation of Section 1, maximal
symmetry breaking is obtained by changing the choice of the vector 
${\bfmath v}$ such that ${\bfmath v}\cdot \bbeta >0$. 
Now there are massive fundamental monopole states associated with the two
positive simple roots $\balpha$ and $\bbeta$. In addition to this, 
the authors of [\Ref{G2},\Ref{W6}] demonstrate the existence of a
threshold bound-state of two fundamental monopoles associated with
the non-simple root ${\balpha}+{\bbeta}$. If we now take the limit
${\bfmath v}\cdot {\bbeta} \rightarrow 0$ the ${\bbeta}$ monopole
becomes massless and the bound-state becomes exactly degenerate
with the ${\balpha}$ monopole. 
Importantly, because the bound-state of [\Ref{G2},\Ref{W6}] 
occurs only in a gauge theory with $N=4$ supersymmetry, the degeneracy
occurs only in this case. Hence the resulting spectrum, in this limit, 
is in complete agreement with the results presented in this paper. 

We should emphasize that the above refers only to the limit of the spectrum
in [\Ref{G2},\Ref{W6}], not of the underlying moduli space. 
In particular the metric on the
moduli space given in [\Ref{G2},\Ref{W6}] is singular in
the limit of non-maximal symmetry breaking. However this divergence
is due to the corresponding zero modes becoming
non-normalizable, which is due in turn to the onset 
of long-range non-abelian fields in this limit. It seems likely that this
singularity could be removed by adding a third monopole, well
separated from the other two, which exactly 
cancels the long-range non-abelian field. 

\appendix{A}

In this appendix we consider in more detail the zero-modes
corresponding to the global non-abelian colour transformations.

Consider first of all the 
Laplace equation \LE\ which must be satisfied by the gauge
transformations in the background of a monopole. 
At distances much greater than the size of the core, it is convenient 
to work in unitary gauge using the
two-patch formalism of Wu and Yang [\Ref{WY}]. In this gauge, in the monopole
background the angular momentum operator is
$$
L_i=-i\epsilon_{ijk}x_jD_k-i{x_i\over r}[{\rm ad}\,t^3(\balpha)].
\efr
The Laplace equation \LE\ can then be expressed as
$$
\left[{1\over r^2}{\partial\,\over\partial
r}\left({r^2{\partial\,\over\partial r}}\right)-{1\over
r^2}L_i^2\right]\Omega=0.
\efr
The solution can be expanded in terms of the eigenvectors of $[{\rm
ad}\,t^3(\balpha)]$: 
$$
\Omega(x_i)=\sum_\bbeta\Omega_\bbeta(x_i)E_\bbeta+\Omega_0(x_i)t^3(\balpha).
\efr
The component associated to $E_\bbeta$ 
can then be expanded in terms of monopole harmonics [\Ref{WY}]
$$
\Omega_\bbeta(r,\theta,\phi)=f(r)Y_{qlm}(\theta,\phi),
\efr
where $q=-\balpha\cdot\bbeta/\balpha^2=\pm1/2$ is the eigenvalue of
$[{\rm ad}\,it^3(\balpha)]$ 
on the generator $E_\bbeta$. The quantum numbers $l$
amd $m$ are the total angular momentum and its $x_3$ component, so that 
$l=0,1/2,1,\ldots$ and $-l\leq m\leq l$. The allowed values of $l$ are $\vert
q\vert,\ \vert q\vert+1,\ldots$. The radial equation is
$$
\left[{1\over r^2}{d\,\over d
r}\left({r^2{d\,\over dr}}\right)-{1\over
r^2}l(l+1)\right]f(r)=0,
\efr
and so $f(r)=r^\alpha$ where $\alpha(\alpha+1)=l(l+1)$. 

For each pair of roots $\pm\bbeta$ with $\bbeta\cdot\balpha=-2\balpha^2/2$,
the solutions which are relevant for the global non-abelian colour
modes are $r^{1/2}Y_{\pm\half\half m}(\theta,\phi)E_{\pm\bbeta}$,
where $m=\pm\half$. Imposing the reality condition
$\Omega^\dagger=-\Omega$, and using the reality properties of the
monopole harmonics
$Y_{qlm}^*(\theta,\phi)=(-)^{q+m}Y_{-ql-m}(\theta,\phi)$ 
we find the solutions 
$$
\eqalign{
\Omega^1(r,\theta,\phi)&=ir^{1/2}\left(Y_{\half\half-\half}(\theta,\phi)
E_{\bbeta}+
Y_{-\half\half\half}(\theta,\phi)E_{-\bbeta}\right),\cr
\Omega^2(r,\theta,\phi)&=r^{1/2}\left(Y_{\half\half-\half}(\theta,\phi)
E_{\bbeta}-
Y_{-\half\half\half}(\theta,\phi)E_{-\bbeta}\right),\cr
\Omega^3(r,\theta,\phi)&=r^{1/2}\left(Y_{\half\half\half}(\theta,\phi)
E_{\bbeta}+ 
Y_{-\half\half-\half}(\theta,\phi)E_{-\bbeta}\right),\cr
\Omega^4(r,\theta,\phi)&=ir^{1/2}\left(Y_{\half\half\half}(\theta,\phi)
E_{\bbeta}-
Y_{-\half\half-\half}(\theta,\phi)E_{-\bbeta}\right).\cr}
\nfr{NMM}

Let us suppose that in the multi-monopole configuration considered in
the text, the monopole is situated along the positive $x_3$ axis
relative to the conglomerate. Near the monopole we can ignore to a
first approximation the
fields of the conglomerate and the global gauge transformations must
satisfy the Laplace equation \LE\ in the back-ground of the monopole alone.
The question to which we now turn is
what linear combinations of the four gauge transformations in \NMM\ 
generate the two global colour rotation modes associated to $E_{\pm\bbeta}$
near the monopole? Each of the monopole harmonics $Y_{\pm\half\pm\half
m}(\theta,\phi)$ vanishes along a particular direction in space and the 
basis of solutions $\Omega^a(x_i)$ has been
chosen to respect the symmetry of the problem.
The first two solutions vanish in the negative $x_3$
direction while the second two vanish in the positive $x_3$
direction. The effect of the fields of the
conglomerate on the first two solutions $\Omega^1(x_i)$ and
$\Omega^2(x_i)$ will be suppressed due to the fact that they
vanish in its direction. The other two solutions become unacceptably
large in the direction of the conglomerate and would not lead to
solutions of the full system. 
For this reason we identify the first two of these 
solutions as generating the two global non-abelian modes.

The explicit expressions for the corresponding zero-modes are
$$\eqalign{
\delta_1W_m&=\left({xt^1(\bbeta)-yt^2(\bbeta)\over2r\sqrt{r+z}},
{yt^1(\bbeta)+xt^2(\bbeta)\over2r\sqrt{r+z}},{\sqrt{r+z}\over2r}t^1(\bbeta),
-{\sqrt{r+z}\over2r}t^2(\bbeta)\right),\cr
\delta_2W_m&=\left({yt^1(\bbeta)+xt^2(\bbeta)\over2r\sqrt{r+z}},
{-xt^1(\bbeta)+yt^2(\bbeta)\over2r\sqrt{r+z}},{\sqrt{r+z}\over2r}t^2(\bbeta),
{\sqrt{r+z}\over2r}t^1(\bbeta)\right),\cr}
\efr
which are valid in the northern hemisphere around the
monopole. These two solutions are related by one of the three almost complex
structures of ${\Bbb R}^4$ namely the one which relates
$x_1\leftrightarrow x_2$ and $x_3\leftrightarrow x_4$. 
Explicitly
$$
J_{mn}=\pmatrix{0&-1&0&0\cr 1&0&0&0\cr 0&0&0&1\cr 0&0&-1&0\cr},
\efr 
and
$$
{\cal J}_{ab}\delta_bW_m=-J_{mn}\delta_aW_n,
\efr
with 
$$
{\cal J}_{ab}=\pmatrix{0&1\cr -1&0\cr}.
\efr

\references

\beginref
\Rref{TP}{G. 't Hooft, Nucl. Phys. {\bf B79} (1976) 276\newline
A.M. Polyakov, JETP Lett. {\bf20} (1974) 194}
\Rref{GNO}{P. Goddard, J. Nuyts and D. Olive, Nucl. Phys. {\bf B125}
(1977) 1}
\Rref{WB2}{E.J. Weinberg, Nucl. Phys. {\bf B203} (1982) 445}
\Rref{WB1}{E.J. Weinberg, Nucl. Phys. {\bf B167} (1980) 500}
\Rref{WB3}{E.J. Weinberg, Phys. Rev. {\bf D20} (1979) 936}
\Rref{WB4}{E.J. Weinberg, Phys. Lett. {\bf B119} 151}
\Rref{A1}{A. Abouelsaood, Phys. Lett. {\bf B137} (1984) 77}
\Rref{A2}{A. Abouelsaood, Nucl. Phys. {\bf B226} (1983) 309}
\Rref{A3}{A. Abouelsaood, Phys. Lett. {\bf B125} (1983) 467}
\Rref{BMMNS1}{A.P. Balachandran, G. Marmo, N. Mukunda, J.S. Nilsson,
E.C.G. Sudarshan and F. Zaccaria, Phys. Rev. {\bf D29} (1984) 2919}
\Rref{G1}{J.P. Gauntlett, Nucl. Phys. {\bf B411} (1994) 443}
\Rref{BAIS}{F.A. Bais, Phys. Rev. {\bf D18} (1978) 1206}
\Rref{NM}{P. Nelson and A. Manohar, Phys. Rev. Lett. {\bf50} (1983) 943}
\Rref{NC}{P. Nelson and S. Coleman, Nucl. Phys. {\bf B237} (1984) 1}
\Rref{MONM}{M.K. Murray, Commun. Math. Phys. {\bf96} (1984) 539;
Commun. Math. Phys. {\bf125} (1989) 661\newline 
M.C. Bowman, Phys. Rev. {\bf D32} (1985) 1569}
\Rref{LVW1}{W. Lerche, C. Vafa and N.P. Warner, Nucl. Phys. {\bf
B324} (1989) 427}
\Rref{OS}{H. Osborn, Phys. Lett. {\bf B83} (1979) 321}
\Rref{JZ}{B. Julia and A. Zee, Phys. Rev. {\bf D11} (1975) 2227}
\Rref{WY}{T.T. Wu and C.N. Yang, Nucl. Phys. {\bf B107} (1976) 365}
\Rref{EW}{F. Englert and P. Windey, Phys. Rev. {\bf D14} (1976) 2728}
\Rref{HS}{J.A. Harvey and A. Strominger, Commun. Math. Phys. {\bf151}
(1993) 221}
\Rref{HEL}{S. Helgason, {\it Differential Geometry, Lie Groups and
Symmetric Spaces\/}, Academic Press 1978} 
\Rref{MO}{C. Montonen and D. Olive, Phys. Lett. {\bf B72} (1977) 117}
\Rref{WL}{R. Ward, Phys. Lett. {\bf B158} (1985) 424\newline
R. Leese, Nucl. Phys. {\bf B344} (1990) 33}
\Rref{SEN}{A. Sen, Phys. Lett. {\bf B329} (1994) 217-221}
\Rref{AG}{L. Alvarez-Gaum\'e, Commun. Math. Phys. {\bf90} (1983)
161\newline
E. Witten, Nucl. Phys. {\bf B202} (1982) 253}
\Rref{BL}{J.D. Blum, Phys. Lett. {\bf B333} (1994) 92}
\Rref{CORN}{J.F. Cornwell, {\it Group Theory in Physics
Vols. 1,2,3\/}, Academic Press 1984}
\Rref{SEI}{N. Seiberg, Nucl. Phys. {\bf B435} (1995) 129}
\Rref{MAN}{ N. Manton, Phys. Lett. {\bf B110} (1982) 54}
\Rref{CAL}{C.J. Callias, Commun. Math. Phy. {\bf 62} (1978) 213}
\Rref{PS}{M.K. Prasad and C.M. Sommerfield, Phys. Rev. Lett. {\bf35}
(1975) 760}
\Rref{NTWO}{A. Klemm, W.Lerche, S. Theisen and S. Yankielowicz,
Phys. Lett. {\bf B344} (1995) 169\newline
P.C. Argyers and A. Farragi, Phys. Rev. Lett. {\bf74} (1995)
3931\newline
P.C. Argyers, M.R. Plesser and A.D. Shapere, Phys. Rev. Lett. {\bf75}
(1995) 1699}
\Rref{N}{P. Nelson, Phys. Rev. Lett. {\bf 50} (1983) 939} 
\Rref{G2}{J.P. Gauntlett and D. A. Lowe, `Dyons and S-duality in 
$N=4$ Supersymmetric Gauge Theory', {\tt hep-th/9601085}}
\Rref{W6}{K. Lee, E.J. Weinberg and P. Yi, `Electromagnetic Duality
and $\SU(3)$ Monopoles', {\tt hep-th/9601097}}
\Rref{EV}{C. Vafa and E. Witten, Nucl. Phys. {\bf B431} (1994) 3; \newline 
J. A. Harvey, G. Moore and A. Strominger, {\tt hep-th/9501096}; \newline 
M. Bershadsky, A. Johansen and C. Vafa, Nucl. Phys. {\bf B448} (1995)
166.}
\Rref{GH}{J.P. Gauntlett and J.A. Harvey, `S duality and the dyon
spectrum in N=2 super Yang-Mills theory', {\tt hep-th/9508156}}  
\Rref{BOTT}{M.R. Bott, Bull. Soc. Math. France, {\bf84} (1956) 251}
\endref
\ciao